\documentclass[twocolumn]{aastex63}

\newcommand{\elsonido}{\textit{El Sonido}}
\newcommand{\herschel}{\textit{Herschel}}
\newcommand{\spitzer}{\textit{Spitzer}}
\newcommand{\hst}{\textit{HST}}
\newcommand{\zphot}{\mbox{$z_\mathrm{phot}$}}

\usepackage{savesym}
\savesymbol{tablenum}
\usepackage[binary-units]{siunitx}
\restoresymbol{SIX}{tablenum}

\renewcommand\micron{\mbox{\si{\micro}{m}}}

\renewcommand\arcdeg{\mbox{$^\circ$}}% 
\renewcommand\arcsec{\mbox{$^{\prime\prime}$}}% 

\newcommand\msun{\mbox{\si{M_\odot}}}
\newcommand\lsun{\mbox{\si{L_\odot}}}
\newcommand\smpy{\mbox{\si{M_\odot.yr^{-1}}}}
\newcommand\mstar{\mbox{$M_\mathrm{star}$}}

\received{May 15, 2021}
\accepted{September 8, 2021}
\submitjournal{ApJ}

\shorttitle{\textit{El Sonido}: Extensive Lensing Survey of Optical and Near-Infrared Dark Objects}
\shortauthors{Sun et al.}
\graphicspath{{./}{figures/}}
\begin{document}

\title{Extensive Lensing Survey of Optical and Near-Infrared Dark Objects (\textit{El Sonido}):\\
\textit{HST} \textit{H}-Faint Galaxies behind 101 Lensing Clusters}

\correspondingauthor{Fengwu Sun}
\email{fengwusun@email.arizona.edu}

\author{Fengwu Sun}
\affiliation{Steward Observatory, University of Arizona, 933 N. Cherry Avenue, Tucson, 85721, USA}

\author{Eiichi Egami}
\affiliation{Steward Observatory, University of Arizona, 933 N. Cherry Avenue, Tucson, 85721, USA}

\author{Pablo G.\ P\'erez-Gonz\'alez}
\affiliation{Centro de Astrobiolog\'ia, Departamento de Astrof\'isica, CSIC-INTA, Cra.\ de Ajalvir km.4 E-28850--Torrej\'on de Ardoz, Madrid, Spain}

\author{Ian Smail}
\affiliation{Centre for Extragalactic Astronomy, Department of Physics, Durham University, South Road, Durham, DH1 3LE, UK}

\author{Karina I. Caputi}
\affiliation{Kapteyn Astronomical Institute, University of Groningen, P.O. Box 800, 9700AV Groningen, The Netherlands}
\affiliation{Cosmic Dawn Center (DAWN), Jagtvej 128, DK2200 Copenhagen N, Denmark}

\author{Franz E. Bauer}
\affiliation{Instituto de Astrofısica, Facultad de Fısica, Pontificia Universidad Catolica de Chile Av. Vicuna Mackenna 4860, 782-0436
Macul, Santiago, Chile}
\affiliation{Millennium Institute of Astrophysics (MAS), Nuncio Monse nor Santero Sanz 100, Providencia, Santiago, Chile}

\author{Timothy D.\ Rawle}
\affiliation{European Space Agency (ESA), ESA Office, Space Telescope Science Institute, 3700 San Martin Drive, Baltimore, MD 21218, USA}

\author{Seiji Fujimoto}
\affiliation{Cosmic Dawn Center (DAWN), Jagtvej 128, DK2200 Copenhagen N, Denmark}
\affiliation{Niels Bohr Institute, University of Copenhagen, Lyngbyvej 2, DK2100 Copenhagen {\O}, Denmark}

\author{Kotaro Kohno}
\affiliation{Institute of Astronomy, Graduate School of Science, The University of Tokyo, 2-21-1 Osawa, Mitaka, Tokyo 181-0015, Japan}
\affiliation{Research Center for the Early Universe, School of Science, The University of Tokyo, 7-3-1 Hongo, Bunkyo-ku, Tokyo 113-0033, Japan}

\author{Ugn{\.{e}} Dudzevi{\v{c}}i{\={u}}t{\.{e}}}
\affiliation{Centre for Extragalactic Astronomy, Department of Physics, Durham University, South Road, Durham, DH1 3LE, UK}

\author{Hakim Atek}
\affiliation{Institut d'Astrophysique de Paris, Sorbonne Universite, CNRS, UMR 7095, 98 bis bd Arago, 75014 Paris, France}

\author{Matteo Bianconi}
\affiliation{School of Physics \& Astronomy, University of Birmingham, Birmingham, B15 2TT, UK}

\author{Scott C.\ Chapman}
\affiliation{{Eureka Scientific, Inc. 2452 Delmer Street Suite 100, Oakland, CA 94602-3017, USA}}
% \affiliation{Department of Physics and Astronomy, University of British Columbia, 6225 Agricultural Road, Vancouver, BC V6T 1Z1, Canada}
% \affiliation{National Research Council, Herzberg Astronomy and Astrophysics, 5071 West Saanich Road, Victoria, BC V9E 2E7, Canada}
% \affiliation{Department of Physics and Atmospheric Science, Dalhousie University, Halifax, NS B3H 4R2, Canada}

\author{Francoise Combes}
\affiliation{Sorbonne Universit\'e, Observatoire de Paris, Universit\'e PSL, CNRS, LERMA, 75014 Paris, France}
\affiliation{Coll\`ege de France, 11 Place Marcelin Berthelot, 75231 Paris, France}

\author{Mathilde Jauzac}
\affiliation{Centre for Extragalactic Astronomy, Durham University, South Road, Durham DH1 3LE, UK}
\affiliation{Institute for Computational Cosmology, Durham University, South Road, Durham DH1 3LE, UK}
\affiliation{Astrophysics Research Centre, University of KwaZulu-Natal, Westville Campus, Durban 4041, South Africa}
\affiliation{School of Mathematics, Statistics \& Computer Science, University of KwaZulu-Natal, Westville Campus, Durban 4041, South Africa}

\author{Jean-Baptiste Jolly}
\affiliation{Department of Space, Earth and Environment, Chalmers University of Technology, Onsala Space Observatory, SE-439 92 Onsala, Sweden}

\author{Anton M. Koekemoer}
\affiliation{Space Telescope Science Institute, 3700 San Martin Drive, Baltimore, MD, 21218, USA}

\author{Georgios E. Magdis}
\affiliation{Cosmic Dawn Center (DAWN), Jagtvej 128, DK2200 Copenhagen N, Denmark}
\affiliation{DTU-Space, Technical University of Denmark, Elektrovej 327, DK-2800 Kgs. Lyngby, Denmark}
\affiliation{Niels Bohr Institute, University of Copenhagen, Lyngbyvej 2, DK2100 Copenhagen {\O}, Denmark}

\author{Giulia Rodighiero}
\affiliation{Dipartimento di Fisica e Astronomia, Universit\`a di Padova, vicolo dell'Osservatorio 3, I-35122 Padova, Italy}
\affiliation{INAF - Osservatorio Astrofisico di Padova, vicolo dell'Osservatorio 5, I-35122 Padova, Italy}

\author{Wiphu Rujopakarn}
\affiliation{Department of Physics, Faculty of Science, Chulalongkorn University, 254 Phayathai Road, Pathumwan, Bangkok 10330, Thailand}
\affiliation{National Astronomical Research Institute of Thailand (Public Organization), Don Kaeo, Mae Rim, Chiang Mai 50180, Thailand}
\affiliation{Kavli Institute for the Physics and Mathematics of the Universe (WPI),The University of Tokyo Institutes for Advanced Study, The University of Tokyo, Kashiwa, Chiba 277-8583, Japan}

\author{Daniel Schaerer}
\affiliation{Observatoire de Gen\`eve, Universit\'e de Gen\`eve, 51, Ch. des Maillettes, 1290 Versoix, Switzerland}

\author{Charles L. Steinhardt}
\affiliation{Cosmic Dawn Center (DAWN), Jagtvej 128, DK2200 Copenhagen N, Denmark}
\affiliation{Niels Bohr Institute, University of Copenhagen, Lyngbyvej 2, DK2100 Copenhagen {\O}, Denmark}

\author{Paul Van der Werf}
\affiliation{Leiden Observatory, Leiden University, P.O. Box 9513, NL-2300 RA Leiden, The Netherlands}

\author{Gregory L. Walth}
\affiliation{The Observatories of the Carnegie Institution for Science, 813 Santa Barbara Street, Pasadena, CA 91101, USA}

\author{John R. Weaver}
\affiliation{Cosmic Dawn Center (DAWN), Jagtvej 128, DK2200 Copenhagen N, Denmark}
\affiliation{Niels Bohr Institute, University of Copenhagen, Lyngbyvej 2, DK2100 Copenhagen {\O}, Denmark}

%% Note that the \and command from previous versions of AASTeX is now
%% depreciated in this version as it is no longer necessary. AASTeX 
%% automatically takes care of all commas and "and"s between authors names.

%% AASTeX 6.3 has the new \collaboration and \nocollaboration commands to
%% provide the collaboration status of a group of authors. These commands 
%% can be used either before or after the list of corresponding authors. The
%% argument for \collaboration is the collaboration identifier. Authors are
%% encouraged to surround collaboration identifiers with ()s. The 
%% \nocollaboration command takes no argument and exists to indicate that
%% the nearby authors are not part of surrounding collaborations.

%% Mark off the abstract in the ``abstract'' environment. 
\begin{abstract}
We present a \spitzer/IRAC survey of $H$-faint ($H_{160} \gtrsim 26.4$, $<$\,5$\sigma$) sources in 101 lensing cluster fields.
Across a CANDELS/Wide-like survey area of $\sim$648\,arcmin$^2$ (effectively $\sim$221\,arcmin$^2$ in the source plane), 
we have securely discovered 53 sources in the IRAC Channel-2 band (CH2, 4.5\,\micron; median CH2$=$22.46$\pm$0.11\,AB mag) that lack robust \hst/WFC3-IR F160W counterparts. 
The most remarkable source in our sample, namely ES-009 in the field of Abell\,2813, is the brightest $H$-faint galaxy at 4.5\,\micron\ known so far ($\mathrm{CH2}=20.48\pm0.03$ AB mag). 
We show that the $H$-faint sources in our sample are massive (median {$M_\mathrm{star} = 10^{10.3\pm 0.3}$\,\msun}), star-forming (median star formation rate {$=100_{-40}^{+60}$\,\smpy}) and dust-obscured ({$A_V=2.6\pm0.3$}) galaxies around a median photometric redshift of {$z=3.9\pm{0.4}$}.
The stellar continua of 14 $H$-faint galaxies can be resolved in the CH2 band, suggesting a median circularized effective radius ($R_\mathrm{e,circ}$; lensing corrected) of $1.9\pm0.2$\,kpc and $<1.5$\,kpc for the resolved and whole samples, respectively.
This is consistent with the sizes of massive unobscured galaxies at $z\sim4$, indicating that $H$-faint galaxies represent the dusty tail of the distribution of a wider galaxy population.
Comparing with the ALMA dust continuum sizes of similar galaxies reported previously, we conclude that the heavy dust obscuration in $H$-faint galaxies is related to the compactness of both stellar and dust continua ($R_\mathrm{e,circ}\sim 1$\,kpc).
These $H$-faint galaxies make up {$16_{-7}^{+13}$}\% of the galaxies in the stellar mass range of {$10^{10}-10^{11.2}$}\,\msun\ at $z=3\sim5$, contributing to $8_{-4}^{+8}$\% of the cosmic star formation rate density in this epoch and likely tracing the early phase of massive galaxy formation.
\end{abstract}

%% Keywords should appear after the \end{abstract} command. 
%% See the online documentation for the full list of available subject
%% keywords and the rules for their use.
%% From the front matter, we move on to the body of the paper.
%% Sections are demarcated by \section and \subsection, respectively.
%% Observe the use of the LaTeX \label
%% command after the \subsection to give a symbolic KEY to the
%% subsection for cross-referencing in a \ref command.
%% You can use LaTeX's \ref and \label commands to keep track of
%% cross-references to sections, equations, tables, and figures.
%% That way, if you change the order of any elements, LaTeX will
%% automatically renumber them.
%%
%% We recommend that authors also use the natbib \citep
%% and \citet commands to identify citations.  The citations are
%% tied to the reference list via symbolic KEYs. The KEY corresponds
%% to the KEY in the \bibitem in the reference list below. 

\section{Introduction} 
\label{sec:01_intro}

% dust obscuration - > SMGs

Dust obscuration is known to play a critical role in reshaping the appearance of star-forming galaxies from the local Universe to the epoch of reionization \citep[e.g.,][]{riechers13,strandet17,marrone18,tamura19,bakx20}.
Significant dust absorption of the rest-frame UV-optical light is an indispensable physical process to produce the thermal continuum radiation seen in the far-infrared (far-IR).
This is typically observed at flux densities of $\gtrsim 1$\,mJy shortward of 1\,mm in the submillimeter galaxies (SMGs) that are discovered in abundance at $z\simeq 1- 4$ \citep[see reviews in][]{casey14,hodge20}.
The high observed dust-to-stellar luminosity ratio and red stellar continuum color of SMGs suggest a high dust attenuation, which is typically $A_V \gtrsim 2$ \citep[e.g.,][]{dacunha15,dudzevic20,dudzevic21}.

% SMGs -> optical/near-IR dark sources
Among all the SMGs discovered routinely since the end of the last century, the optically faint SMG population has been of particular interest \citep[e.g.,][]{smail99,dey99,bertoldi00,frayer00,frayer04,dannerbauer02,wang07,tamura10}.
Identified at millimeter wavelengths with large IR luminosities ($L_\mathrm{IR}\gtrsim 10^{12}$\,\lsun), these galaxies are found to be faint or even undetected in the optical/near-IR, suggesting heavily reddened stellar continua due to strong dust obscuration and/or high redshift ($z\gtrsim 3$).
This was highlighted by the study of HDF850.1, the brightest 850\,\micron\ source discovered in the \textit{Hubble} Deep Field \citep{hughes98}, which was confirmed later to be an optical/near-IR-faint galaxy ($I\gtrsim 29$, $K\sim23.5$, \citealt{dunlop04,cowie09}) at $z=5.18$ with millimeter interferometry \citep{walter12}.
Later ALMA continuum observations suggested that 15$\sim$20\% of SMGs remain undetected in deep ground-based near-IR images ($K>24.4$, \citealt{simpson14}; $K>25.3$, \citealt{dudzevic20,smail20}). 
Similar percentages of optical/near-IR-dark SMGs were also presented by studies based on deep \hst/WFC3-IR F160W data ($H_{160}>27$; \citealt{chen15}, \citealt{franco18}), and these galaxies are often referred to as \hst-dark, $H$-dropout or $H$-faint galaxies \citep[e.g.,][]{yamaguchi19,wang19}.

% Implication of near-IR dark galaxies
These optical/near-IR-faint SMGs are likely highly dust-obscured analogs of local ultraluminous infrared galaxies (ULIRGs, $L_\mathrm{IR} \simeq 10^{12} - 10^{13}$\,\lsun), which can be undetectable at high redshift even with deep optical/near-IR imaging \citep[e.g.,][]{dey99,chapman01,chapman02,frayer04,smail20}.
This implies that the current mass-selected galaxy sample at $z \gtrsim 4$ (de facto based on observed $H/K$-band photometric catalogs) may miss a substantial fraction of dusty galaxies at the most massive end ($M_\mathrm{star}\gtrsim 10^{11}$\,\msun; e.g., \citealt{caputi15,wang16,alcalde19}).
These sources can contribute to $\sim 10$\% of the cosmic star-formation rate density (CSFRD) at $z\simeq 3-5$ \citep{wang19,yamaguchi19,williams19,dudzevic20,bouwens20}, and thus the latest ALMA-based CSFRD can be higher than the previous UV/optical estimates in this redshift range \citep{gruppioni20}. 
Furthermore, the red stellar continuum colors of these optical/near-IR-faint SMGs make them natural contaminants of quiescent/post-starburst galaxy samples at $z\simeq 3-4$ \citep[e.g.,][]{simpson17,schreiber18}.

% near-IR-dark galaxies - > selection using H - 4.5
An efficient selection of optical/near-IR-faint SMGs can also be initiated from IR imaging surveys with deep $H/K$-band and \spitzer/IRAC \citep{irac} coverage \citep[e.g.,][]{ivison04,rodighiero07,huang11,wang12,caputi12,caputi14,wang16,alcalde19}.
These IRAC-selected extremely red objects (EROs) are the higher-redshift extension of optical/near-IR-selected ones  \citep[e.g.,][]{smail02,mccarthy04}, which consist of both dusty star-forming and evolved passive systems.
As highlighted in \citet{wang19}, 39 out of the 63 $H$-faint galaxies ($H_{160}\gtrsim 27$ and $\mathrm{CH2} < 24$) in the CANDELS fields {\citep{koekemoer12,grogin11}} were found to be brighter than 0.6\,mJy at 870\,\micron, suggesting a typical SFR of $\sim 310$\,\smpy.
These ALMA-detected $H$-faint galaxies, found in abundance ($\sim530$\,deg$^{-2}$ by \citealt{wang19}), are proposed as a critical tracer of the early phase of massive galaxy formation history. 
However, most of their physical properties still remain highly uncertain except for rare cases with spectroscopic confirmations \citep[e.g.,][]{walter12,wang19,zhou20,umenhata20,caputi21,mitsuhashi21} due to their non-detections and faintness at most wavelengths.

% Lensing Field -> advantage
In this work, we conduct an Extensive Lensing Survey of Optical/Near-IR Dark Objects (\textit{El Sonido}). 
This is a \spitzer/IRAC survey of $H$-faint galaxies in 101 lensing cluster fields with archival \hst/WFC3-IR F160W data at a median $5\sigma$ depth of $H_{160} \sim 26.4$.
Across a CANDELS/Wide-like total survey area of $\sim$648\,arcmin$^2$ (effectively $\sim$221\,arcmin$^2$ in the source plane), the depth of this dataset is also CANDELS-like ($H_{160} > 27$) assuming a typical lensing magnification of $\mu = 2$.
We identify 53 $H$-faint galaxies that are robustly detected at S/N$>$5 in the IRAC/CH2 band (median CH2=$22.46\pm0.11$) but without a significant counterpart in the F160W band (S/N$<$5).
This yields a \citet{wang19}-like sample but $\sim2$ times brighter owing to the lensing magnification, facilitating further imaging and spectroscopic observations with ALMA and \textit{JWST}.
Moreover, lensing magnification also allows us to spatially resolve the less obscured stellar continua of 14 strongly magnified $H$-faint galaxies with \spitzer/IRAC at 4.5\,\micron. 
A comparable study for the unlensed sample will not be possible until the operation of \textit{JWST}.

% This work
This paper is arranged as follows:
Section~\ref{sec:02_data} introduces the massive galaxy cluster sample and all the utilized \hst, \spitzer, and \herschel\ data as well as the corresponding data reduction procedures.
Section~\ref{sec:03_phot} describes the fundamental measurements with our data, including the source selection, photometry, surface brightness profile modeling and stacking analysis.
In Section~\ref{sec:04_ana} we present and discuss the physical properties of $H$-faint galaxies.
The conclusions can be found in Section~\ref{sec:05_sum}.
Throughout this paper, we assume a flat $\Lambda$CDM cosmology ($h= 0.7$, $\Omega_m = 0.3$) and a \citet{chabrier03} initial mass function. 
The AB magnitude system \citep{abmag} is used to express source brightnesses in the near/mid-IR.

\section{Data} 
\label{sec:02_data}

\subsection{The Cluster Sample}
\label{ss:02a_clu}

We have selected 101 lensing cluster fields with sufficient imaging data to search for $H$-faint galaxies.
This sample includes clusters from four subsets: (\romannumeral 1) six \hst\ Frontier Field clusters (HFF, \citealt{lotz17}; same as BUFFALO, \citealt{steinhardt20}), 
(\romannumeral 2) 21 CLASH clusters (\citealt{postman12}), (\romannumeral 3) 41 RELICS clusters \citep{coe19}, 
and (\romannumeral 4) 
33 additional clusters observed by the two \herschel\ Key Programs, the \herschel\ Lensing Survey (HLS; \citealt{egami10}, \citealt{sun21}) and the Local Cluster Substructure Survey (LoCuSS; \citealt{smith10}) with archival \hst/WFC3-IR F160W imaging data on MAST\footnote{Mikulski Archive for Space Telescopes (MAST),
\href{https://archive.stsci.edu/}{https://archive.stsci.edu/}} that were publicly available as of November 2020.
All of these clusters have been observed by \hst/WFC3-IR in the F160W band and \spitzer/IRAC in the CH1/CH2 at various depths, as further discussed in Appendix \ref{apd:depth}.

\subsection{WFC3-IR and IRAC Data}
\label{ss:02b_data}

We briefly summarize the utilized \hst/WFC3-IR F160W, \spitzer/IRAC CH1 (3.6\,\micron) and CH2 (4.5\,\micron) data of the 101 lensing cluster fields in Table~\ref{tab:01_cluster}, including the observation program IDs, total archival scientific integration time ($t_\mathrm{obs}$) and $5\sigma$ depths.

%%% HST data and reduction

WFC3-IR F160W data are taken from 67 \hst\ observation programs. 
We obtained a uniform reduction of the data with a standard \textsc{drizzlepac} v3.1.8 \citep{2012drzp} routine.
Our data processing started from the calibrated, flat-fielded individual exposures (``\textsc{\_flt}" images).
If multiple observation sessions were found, we would compute and correct the internal astrometric offset between each individual exposures based on the archival drizzled products (``\textsc{\_drz}" images).
We adopted a \textsc{pixfrac} parameter of 0.8 and an output pixel size of 0\farcs06\,pixel$^{-1}$.
For the simplicity of data processing in the six HFF clusters, we only used the data taken by the BUFFALO program (GO 15117, PI: Steinhardt; \citealt{steinhardt20}) because of ({\romannumeral 1}) a larger sky coverage than the original HFF data \citep{lotz17}, and ({\romannumeral 2}) a nearly consistent depth of F160W data as those of the other 95 fields.
If an $H$-faint source lies in the area that HFF covered, we use the image products\footnote{\href{https://archive.stsci.edu/prepds/frontier/}{https://archive.stsci.edu/prepds/frontier/}} processed by the HFF team.

%%% IRAC data and reduction
IRAC CH1 and CH2 data were taken from 68 \spitzer\ observing programs. 
For simplicity, we directly used the reduced \spitzer\ Frontier Field data \citep{lotz17} for the six HFF clusters. 
In the remaining 95 cluster fields where the IRAC data are generally much shallower than those of the HFF, we reduced the data uniformly using a standard \textsc{mopex} routine. 
Our IRAC data processing started from the archival level 1 (BCD) products, and the output pixel size was set as 0\farcs6\,pixel$^{-1}$.

%%% Astrometry calibration
The output frames of our WFC3-IR and IRAC image products were registered with the \textit{Gaia} DR2 \citep{gaiadr2}.
We first extracted source catalogs in the IRAC bands with \textsc{SExtractor} \citep{sex} and cross-matched with the \textit{Gaia} catalog {(typically 20--30 stars)} to correct the world coordinate system (WCS) offsets. 
This achieved a final astrometric error of $\lesssim 0\farcs1$ in IRAC CH1/2. 
Because the field of view (FoV) of WFC3-IR is only 4.6\,arcmin$^2$, much smaller than the area of IRAC image products, it is very likely that the number of \textit{Gaia} stars falling in the WFC3-IR coverage is not large enough for a reliable astrometric registration. 
Therefore, we registered the output frames of WFC3-IR images using the astrometry-corrected IRAC CH1 images with a similar source extraction and cross-matching pipeline. 
The absolute astrometric error in WFC3-IR images was comparable to that of the IRAC images as a consequence.

\subsection{Ancillary Data}
\label{ss:02c_anci}

In order to further characterize the $H$-faint sample, we include other ancillary data for the 32 cluster fields in which we have identified $H$-faint sources (see Section~\ref{sec:03_phot}).
The final constructed dataset in these fields therefore consists of data in 15 bands, namely 
(1) five \hst\ bands: WFC3-IR F105W, F110W, F125W, F140W and F160W, 
(2) five \spitzer\ bands: IRAC CH1, CH2, CH3 (5.8\,\micron), CH4 (8.0\micron) and MIPS 24\,\micron\ \citep{mips}, and 
(3) five \herschel\ bands: PACS 100, 160\,\micron\ \citep{pacs}, SPIRE 250, 350 and 500\,\micron\ \citep{spire}. 

\hst\ data--- Based on an archival search, we have found WFC3-IR/F105W data in 30 out of the 32 clusters, F110W in 8, F125W in 28 and F140W in 22.
The data reduction routine is the same as described in Section~\ref{ss:02b_data}.
Note that we only processed the BUFFALO data (F105W and F125W) for the HFF clusters. 
For sources that are within the coverage of the original HFF data, we directly used the F105W, F125W and F140W data reduced by the HFF team.
The expected depth of \hst/ACS F814W data is comparable to that observed in the WFC3-IR F105W band.
Because of the non-detection of the stacked sources in F105W band (Section~\ref{ss:03d_stack}) and the heavily reddened stellar continuum, we do not expect F814W and bluer \hst\ data to provide any useful spectral energy distribution (SED) constraint.
Therefore, we did not include any \hst/ACS data for the analysis.

\spitzer\ data--- We identified archival IRAC CH3 and CH4 data for 10 out of the 32 cluster fields with $H$-faint objects, and the data were processed using the same method as described in Section~\ref{ss:02b_data}. 
We also included MIPS 24\,\micron\ data for 13 cluster fields. 
MIPS data reduction started from the archival post-BCD products, and we used \textsc{mopex} v1.8 for the flat fielding, artifact removal and image mosaicking. 
The output pixel size is 1\farcs2\,pixel$^{-1}$, and the typical beam size is 6\arcsec.

\herschel\ data--- \herschel/PACS data at 100 and 160\,\micron\ are available for 17 cluster fields.
16 of these clusters were observed by the HLS \citep{egami10} and A370 was observed by the PACS Evolutionary Probe \citep[PEP;][]{lutz11}. 
The observational settings and reduction of PACS data were detailed in \citet{rawle16} and \citet{sun21} and the output pixel scale is 1\arcsec\,pixel$^{-1}$ at 100\,\micron\ and 2\arcsec\,pixel$^{-1}$ at 160\,\micron.
The typical resolution of PACS data is 7\arcsec\ at 100\,\micron\ and 12\arcsec\ at 160\,\micron.
\herschel/SPIRE data at 250, 350 and 500\,\micron\ are available for all the 32 cluster fields. 
Among them, 16 cluster fields were observed by the HLS in the ``deep'' mode to the confusion-limit depth (RMS$\sim$6\,mJy/beam at 250\,\micron), and 14 cluster fields were observed by the HLS in the ``snapshot'' mode (RMS$\sim$10\,mJy/beam at 250\,\micron). 
The remaining two clusters, A370 and CLJ0152.7-1357, were observed as parts of the \herschel\ Multi-tiered Extragalactic Survey \citep[HerMES;][]{oliver12} to the confusion-limit depth.
The observational settings of the HLS clusters and the data reduction procedure were described in \citet{rawle16} and \citet{sun21}. 
The output pixel sizes are 6, 9, 12\arcsec\ at 250--500\,\micron, which are about 1/3 of the beam sizes in the corresponding bands.

\section{Source Extraction and Measurements} \label{sec:03_phot}

\subsection{Source Extraction and Photometry}
\label{ss:03a_source}
\begin{figure*}[!ht]
\centering
\includegraphics[width=\linewidth]{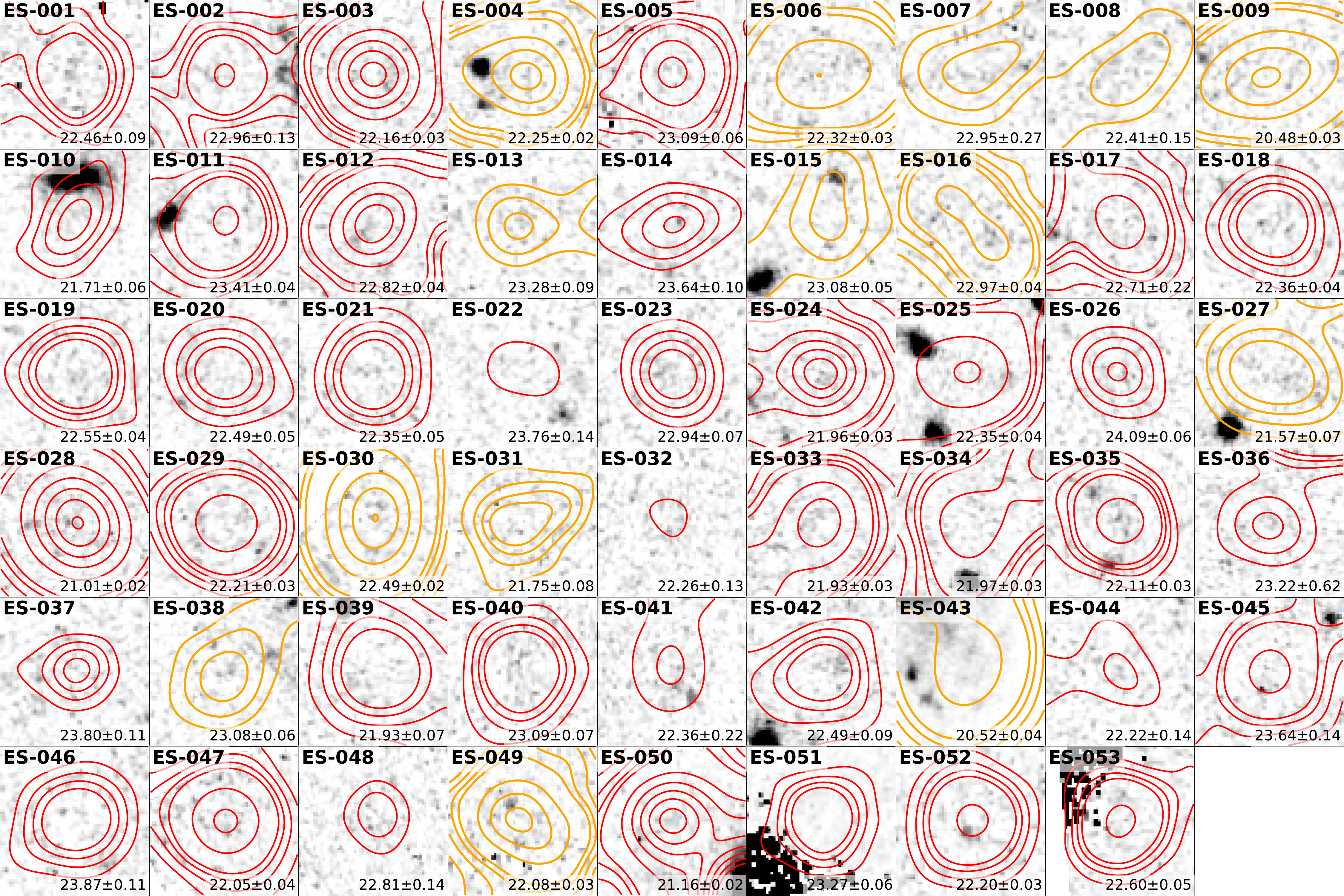}
\caption{Cutout images of all the 53 $H$-faint sources presented in this work. 
\hst/WFC3-IR F160W images are shown in the background, and \spitzer/IRAC CH2 (4.5\,\micron) images are shown as red (if unresolved) or orange (if resolved) contours.
Contour levels are 4, 6, 8, 10, 20, 30$\sigma$ ..., from the outside in.
The IRAC CH2 magnitude of each source is noted at the lower-right corner of each image.
The size of each cutout image is 3\farcs6$\times$3\farcs6.
}
\label{fig:all_sources}
\end{figure*}

To construct a sample of gravitationally lensed $H$-faint galaxies in all 101 cluster fields, we first extracted sources with \textsc{SExtractor} v2.19.5 in both WFC3-IR/F160W and IRAC/CH2 maps.
We extracted sources in the CH2 maps instead of the CH1 maps because $H$-faint sources generally show red IRAC colors and thus the CH2 S/N is higher than that in the CH1.
We fine-tuned the \textsc{SExtractor} configuration files to request five continuous pixels above 1.8$\sigma$ for detections in the CH2 band, and ten continuous pixels above $2\sigma$ in the F160W band. 
We also applied an aggressive set of background subtraction (\textsc{back\_filtersize}=3, \textsc{back\_size}=24) and deblending parameters (\textsc{deblend\_mincont}=$10^{-6}$) to obtain better source deblending and extraction in crowded fields.

We then cross-matched the extracted F160W and CH2 catalogs to select CH2 sources that ({\romannumeral 1}) are within the F160W coverage and $>$1\arcsec\ away from the edge, and ({\romannumeral 2}) have no matched F160W counterpart within a maximum separation of 1\arcsec\ or the brightness of the matched counterpart (measured by \textsc{SExtractor} as \textsc{mag\_auto}) is fainter than that of the CH2 by at least three magnitudes.
We further purified this sample by conducting aperture photometry at the centroids of CH2 sources in the F160W, CH1 and CH2 bands.
We adopted fixed aperture sizes of $r = 0\farcs4$ in F160W and 1\farcs8 in CH1/2, and aperture correction factors are computed from the corresponding 2D point-spread functions (PSFs; 1.20, 1.42, 1.46 times respectively). 
{We note that the point-source assumption for aperture correction in the F160W band may not be valid because a sample of $H$-faint sources can be resolved even with IRAC (Section~\ref{ss:03b_galfit}).
Assuming exponential source profile with circularized effective radius $R_\mathrm{e,circ}=R_\mathrm{e.maj}\sqrt{b/a}=0\farcs45$ and axis ratio $b/a=0.2$ for IRAC-resolved sources, and $R_\mathrm{e,circ}=0\farcs25$, $b/a=0.4$ for IRAC-unresolved sources, additional aperture correction factors of 3.15 and 1.57 times would be needed for both scenarios, respectively.
Such types of aperture correction factors are later applied for the photometry of bright extended sources like ES-009 and stacked sources in Section~\ref{ss:03d_stack}.
}
The sky background was subtracted using the median of sigma-clipped local annulus, and a photometric uncertainty was computed using the RMS of that.
We removed all 110 CH2 sources with F160W counterparts detected above $5\sigma$.

After another round of visual inspection, we further removed 38 spurious sources that were clearly detected in the F160W band but blended in CH2, and four marginally detected CH2 sources without any counterpart in the CH1 band. 
Therefore, our final $H$-faint source sample consists of 53 sources detected in 32 lensing cluster fields that are ({\romannumeral 1}) undetected in the WFC3-IR/F160W band ($<5\sigma$, typically $H_{160}\gtrsim 26.4$ {assuming point-source model}), ({\romannumeral 2}) robustly detected in the IRAC CH2 band at $>5\sigma$ (median CH2 magnitude of CH2=$22.46\pm0.11$), and ({\romannumeral 3}) showing a counterpart in the IRAC CH1 band through visual inspection (median CH1 magnitude of CH1=$23.00\pm0.11$).
We assigned IDs for these sources from ES-001 to ES-053 by the increasing order of R.A.\ (ES\,$=$\,El Sonido).
Figure~\ref{fig:all_sources} shows the cutout images of all 53 $H$-faint sources in our sample. 
The coordinates and photometric measurements are presented in Table~\ref{tab:02_phot}.

{
We also investigate the source detectability and photometry accuracy in the F160W band with a larger aperture size ($r_\mathrm{aper}=0\farcs6$). 
With this increased aperture, three sources are detected above a 5$\sigma$ detection threshold (ES-008, ES-041, and ES-042).  
In two cases (ES-041 and ES-042), the detections are caused by faint clumps offset from the IRAC centroids by 0\farcs5, which may or may not be associated with the IRAC-detected galaxies.  
In the case of ES-008, we clearly see a faint extended source, whose significance of detection has increased with a 0\farcs6 aperture.  
We have also confirmed that 70\% of the sources remain undetected even if we lower the detection threshold from 5 to 3$\sigma$.  
These indicate that our sample of $H$-faint galaxies is sufficiently robust against the choice of photometric aperture and detection threshold.
}

\begin{figure}[!t]
\centering
\includegraphics[width=\linewidth]{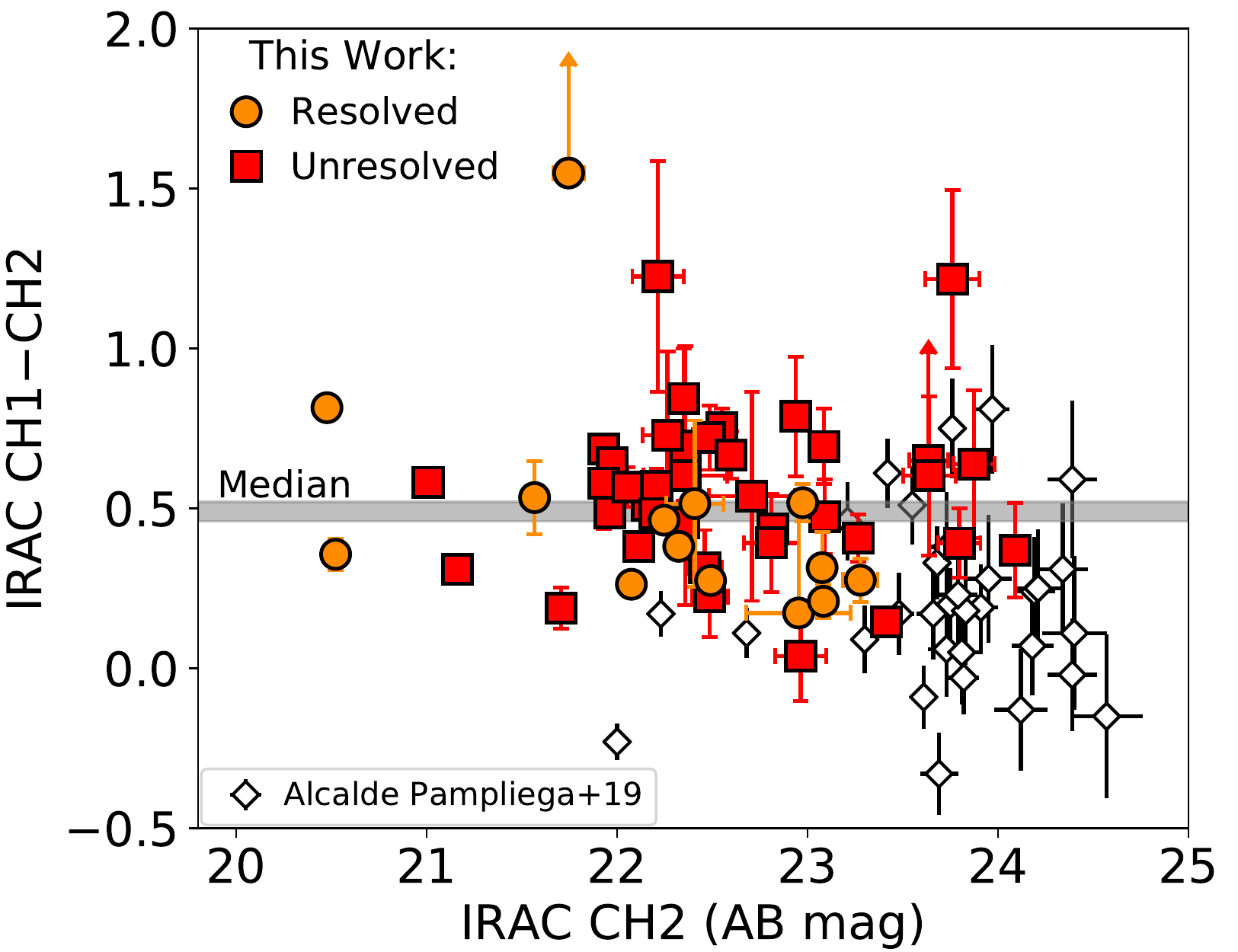}
\caption{IRAC CH1$-$CH2 color versus observed CH2 magnitude for resolved sources (orange circles) and unresolved sources (red squares).
$H$-faint galaxies selected in the CANDELS/GOODS fields ($H_{160}\gtrsim27.3$; \citealt{alcalde19}) are shown as empty diamonds for comparison.
The median IRAC CH1--CH2 color ($0.49\pm0.03$) is indicated by the horizontal grey solid line.
}
\label{fig:cmd}
\end{figure}

Owing to lensing magnification, sources in this sample are brighter than the $H$-faint galaxies in \citet{wang19} and \citet{alcalde19} by $\sim$0.9 and 1.3\,mag at 4.5\,\micron, respectively.
All of the sources show red $H_{160}-\mathrm{CH2}$ colors of $>$2.5 (median $H_{160}-\mathrm{CH2} > 3.9$), consistent with those of the $H$-faint sources selected in CANDELS field \citep{caputi12,wang19}. 
{We note that such $H_{160}-\mathrm{CH2}$ colors are calculated assuming point-source model, and extended source model assumption ($R_\mathrm{e,circ}\simeq 0\farcs25 - 0\farcs45$) will lead to bluer colors by $\sim0.5$\,mag.}
As shown in Figure~\ref{fig:cmd}, the median IRAC CH1--CH2 color is $0.49\pm 0.03$, suggesting a reddened stellar continuum in the IRAC bands or the presence of strong emission lines \citep{alcalde19}.
Given the $H-\mathrm{CH2}$ colors, brown dwarfs with a spectral type cooler than T7 may contaminate our sample \citep[e.g.,][]{kirkpatrick21}. 
However, the $\mathrm{CH1}-\mathrm{CH2}$ colors of these brown dwarfs are very red ($>$1.2),  and the only source with such a secure red $\mathrm{CH1}-\mathrm{CH2}$ color in our sample (ES-031) can be resolved with IRAC CH2. 
Two sources (ES-022/44) tentatively fall in such a red $\mathrm{CH1}-\mathrm{CH2}$ color range of brown dwarf, and we cannot rule out the possibility of such a contamination.
However, the derived physical properties of overall sample will not change even if these sources are excluded.

Among 18 sources for which the IRAC CH3/CH4 data exist, we find that seven (39$\pm$11\%) and eight (44$\pm$12\%) sources can be detected at S/N$>$4 at 5.8\,\micron\ and 8.0\,\micron\ using an $r=2\farcs4$ aperture. 
We further measured the MIPS 24\,\micron\ flux densities of 21 sources for which the data exist. 
Among them, five sources can be extracted at S/N$>$4 using $r=3\farcs5$ aperture with an aperture correction factor of 2.80, further discussed in Section \ref{ss:03c_notable}.
A similar fraction of MIPS-detected sources is also reported for the $H$-faint galaxy sample in \citet{alcalde19}.
The typical $4\sigma$ upper limit of flux densities at 24\,\micron\ is 61\,\si{\micro Jy}.

For 30 sources with existing PACS data, we obtained photometric measurements with $r=4\arcsec$ and 8\arcsec\ apertures at 100 and 160\,\micron.  ES-009, ES-027, and ES-028 are the only sources extracted at S/N$>$4 in at least one band.
The typical $4\sigma$ upper limit of flux densities is 2\,mJy at 100\,\micron\ and 4\,mJy at 160\,\micron.

SPIRE 250-500\,\micron\ flux densities of 21 sources with MIPS data were extracted using the 24\,\micron\ priors with \textsc{xid+} \citep{hurley17}, and ES-009 is the only source detected at $>4\sigma$ in any SPIRE band, as further discussed in Section \ref{ss:03c_notable}. 
The typical $4\sigma$ upper limit of SPIRE flux densities is 8--9\,mJy in all the three bands.

\subsection{Profile Modeling}
\label{ss:03b_galfit}

We further modeled the IRAC/CH2 surface brightness profiles of all the 53 sources using \textsc{galfit} \citep{galfit}.
We only modeled their profiles in the CH2 maps instead of the CH1 maps because ({\romannumeral 1}) the full widths at half maximum (FWHMs) of IRAC PSFs in CH1 and CH2 are similar, and ({\romannumeral 2}) the S/N in CH2 is higher than that in CH1.
\citet{sun21} demonstrate that IRAC can resolve the lensed stellar continua of dusty star-forming galaxies at $z\sim 2$ in the rest-frame $J$ and $H$ bands.
At a source redshift of $z = 4$ and a lensing magnification factor of $\mu=2$, IRAC would be able to resolve a source with a circularized effective radius greater than $\sim 2$\,kpc.

We first constructed the CH2 PSF models from the stars in the mosaic images of all the cluster fields using a \textsc{photutils} routine \citep{photutils}. 
These PSF kernels were then used by \textsc{galfit} for model convolution.
We then extracted a catalog of neighboring field sources from F160W images, and the positions and profile parameters (semi-major axis, axis ratio, positional angle) were then used as prior information for the profile modeling in CH2.
We tried both the S\'{e}rsic and point-source models for the $H$-faint sources, and the final adopted models represent fittings with a higher quality (lower $\chi^2$ or more reasonable parameters, e.g., S\'ersic index $n \lesssim 4$).

\begin{figure}[!t]
\centering
\includegraphics[width=\linewidth]{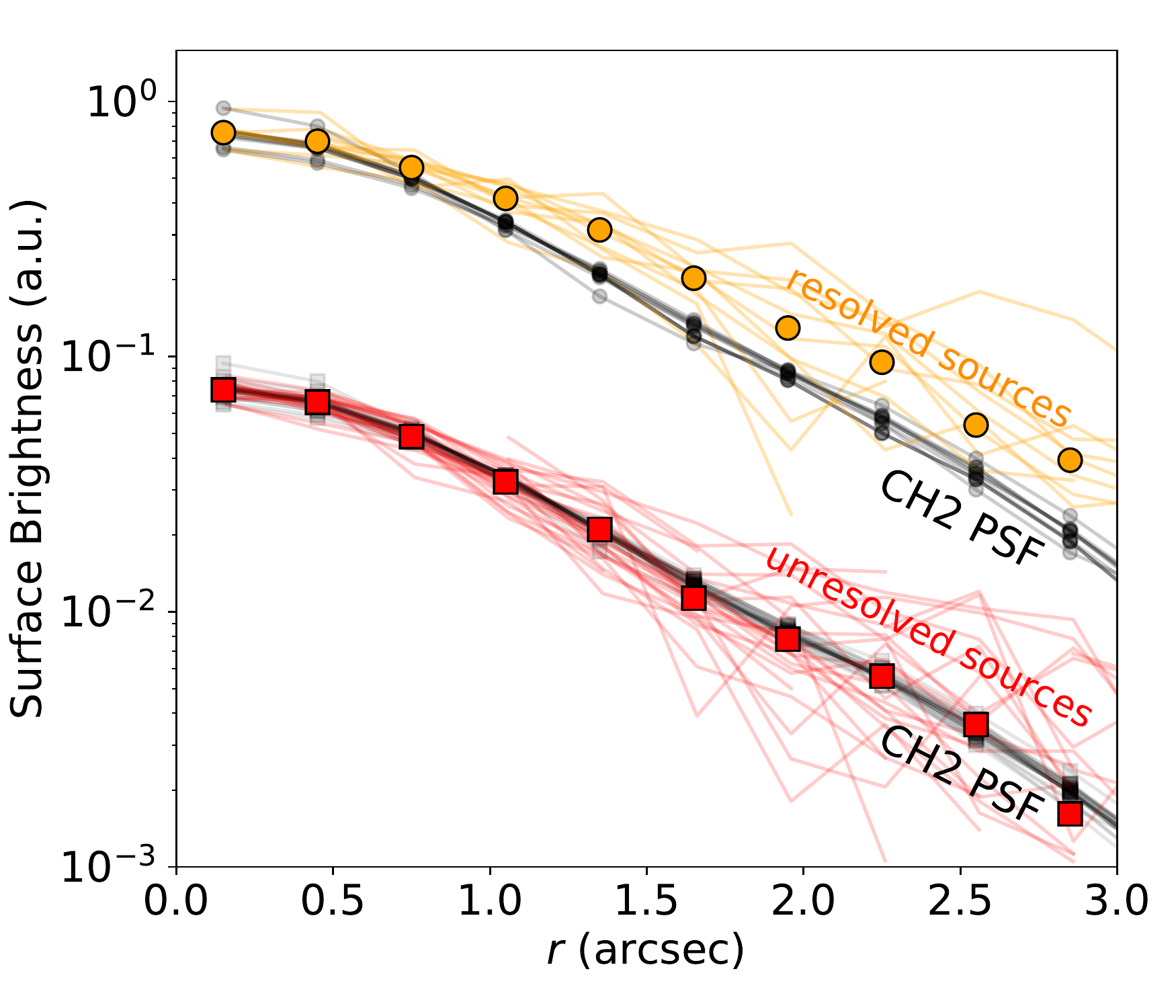}
\vspace{-7mm}
\caption{Radial surface brightness profiles of resolved (top; orange lines) and unresolved sources (bottom; red lines) in the IRAC CH2 band in arbitrary units (a.u.). 
All profiles are normalized by the peak brightness derived with a Gaussian fit, but we displace the two groups vertically by a factor of 10 for clarity.
Radial profiles of the IRAC CH2 PSFs, constructed from the same fields of the $H$-faint sources, are shown as reference in black lines.
The stacked radial profiles of resolved and unresolved sources are indicated by orange circles and red squares, respectively.
}
\label{fig:cog}
\end{figure}

14 out of the 53 $H$-faint sources are spatially resolved in the IRAC/CH2 band.
The radial surface brightness profiles of resolved and unresolved sources are displayed in Figure~\ref{fig:cog} separately.
We derived a mean S\'ersic index of $n_\mathrm{mean} = 0.8 \pm 0.1$ {(typical uncertainty for individual sources is 0.6)}. 
This is consistent with the average S\'ersic index ($n \sim 1$) of the stellar continua of SMGs at $z\sim 2$ measured in either the F160W band \citep[unlensed sample; e.g.,][]{chen15,fujimoto18,lang19} or IRAC bands \citep[lensed sample;][]{sun21}.
Most of the resolved sources show an elongated shape in the IRAC CH2 (mean axis ratio $b/a=0.20\pm0.05$), reflecting a strongly lensed nature or presence of a multiple-component structure.
The \textsc{galfit}-derived magnitudes are also consistent with the aperture-photometry ones (median difference is $0.04\pm0.03$\,mag), suggesting that the source blending will only have a limited influence on the aperture-photometry measurements.

For the resolved sources, we report their circularized effective radii {($R_\mathrm{e, circ}$)} in Table~\ref{tab:02_phot}.
The mean $R_\mathrm{e, circ}$ of the 14 resolved sources is $0\farcs46\pm0\farcs04$. % (corresponding to a physical scale of 3.2\,kpc at $z=4$ before lensing correction).
We also derived the $R_\mathrm{e, circ}$ using the 1D radial surface brightness profiles of these 14 sources. 
Through a Gaussian fitting of the radial profiles of $H$-faint sources and corresponding PSF models, we deconvolved the Gaussian RMS widths of the sources by those of the PSFs and derived the circularized effective radii $R_\mathrm{e, circ}^\mathrm{1D}$. 
The mean ratio between the effective radii measured on 2D maps (\textsc{galfit}) and 1D profiles is $R_\mathrm{e, circ} / R_\mathrm{e, circ}^\mathrm{1D} = 0.9 \pm 0.1$, suggesting a good consistency.

\subsection{Notable Sources}
\label{ss:03c_notable}

\begin{figure*}[!ht]
\centering
\includegraphics[width=\linewidth]{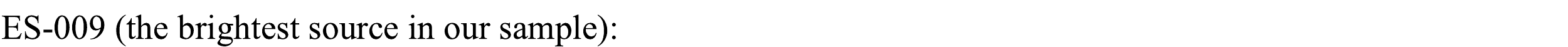}
\includegraphics[width=0.55\linewidth]{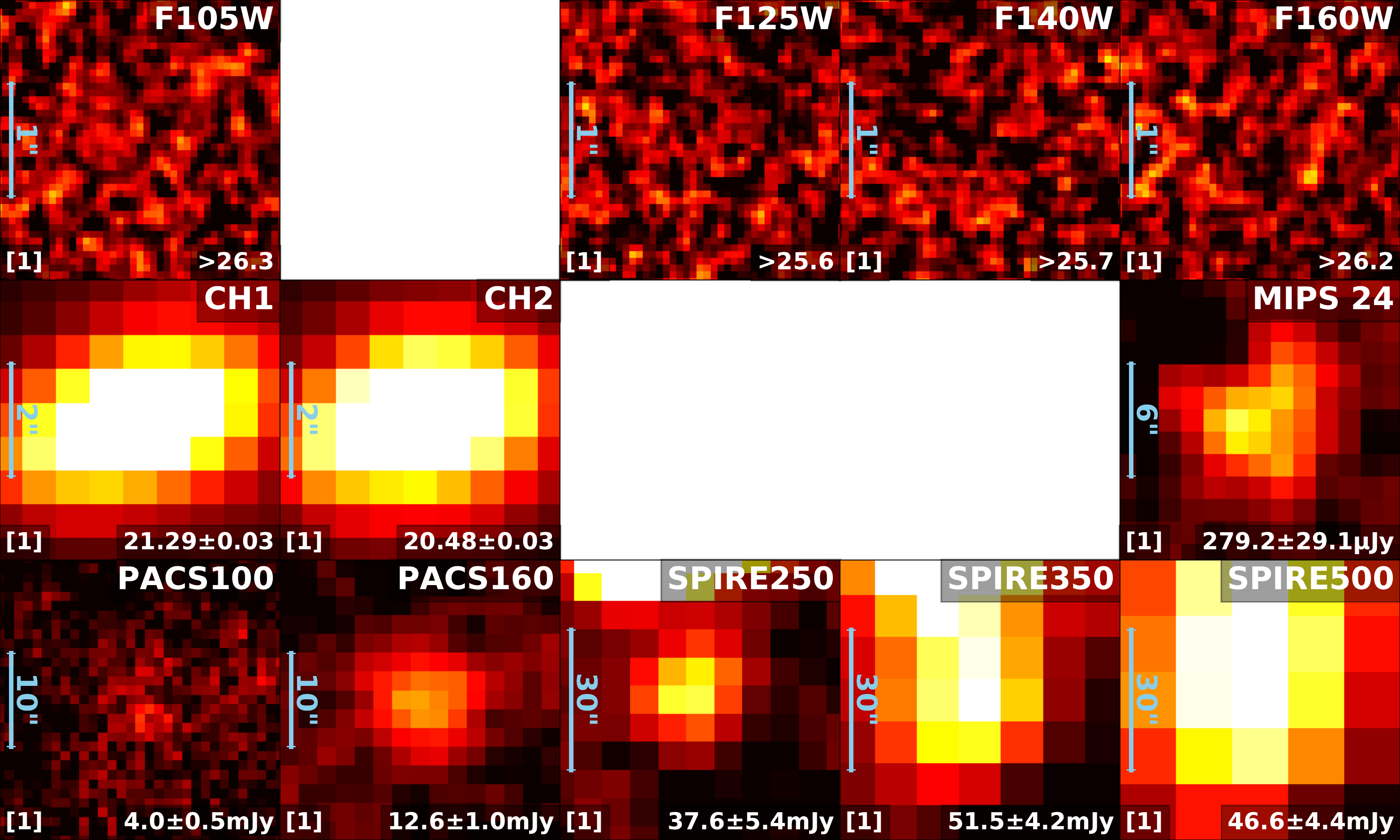}
\includegraphics[width=0.44\linewidth]{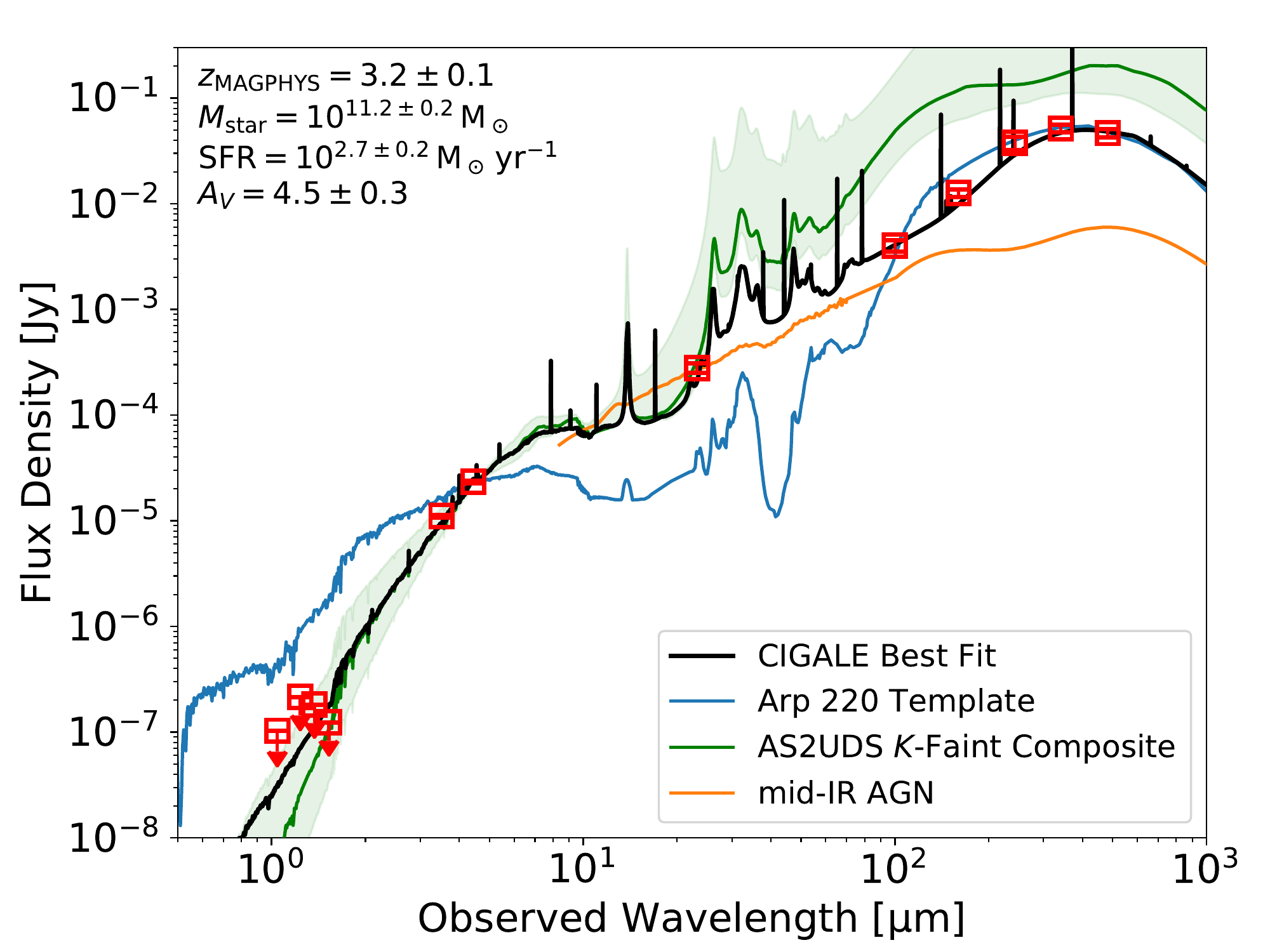}
\includegraphics[width=\linewidth]{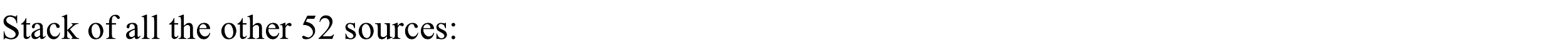}
\includegraphics[width=0.55\linewidth]{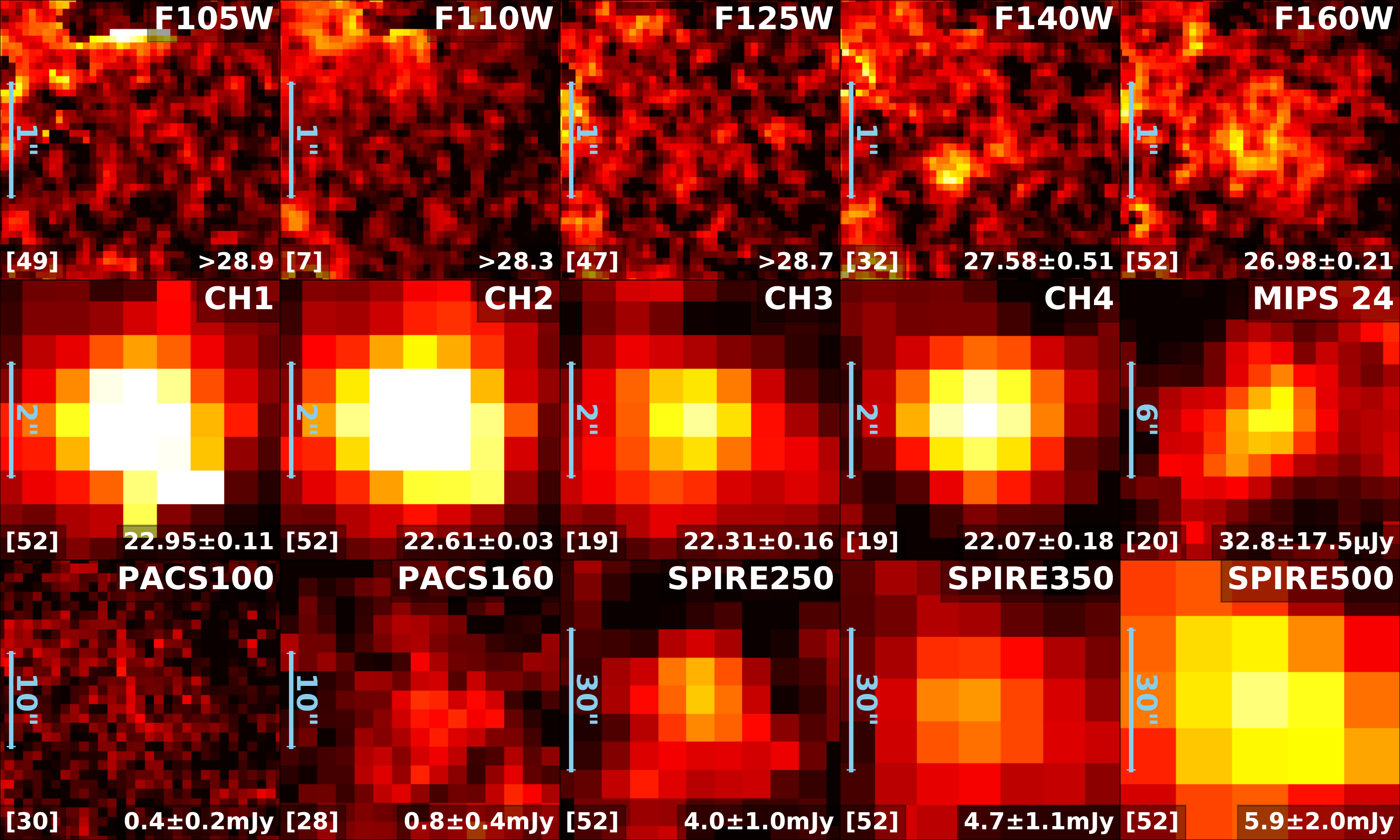}
\includegraphics[width=0.44\linewidth]{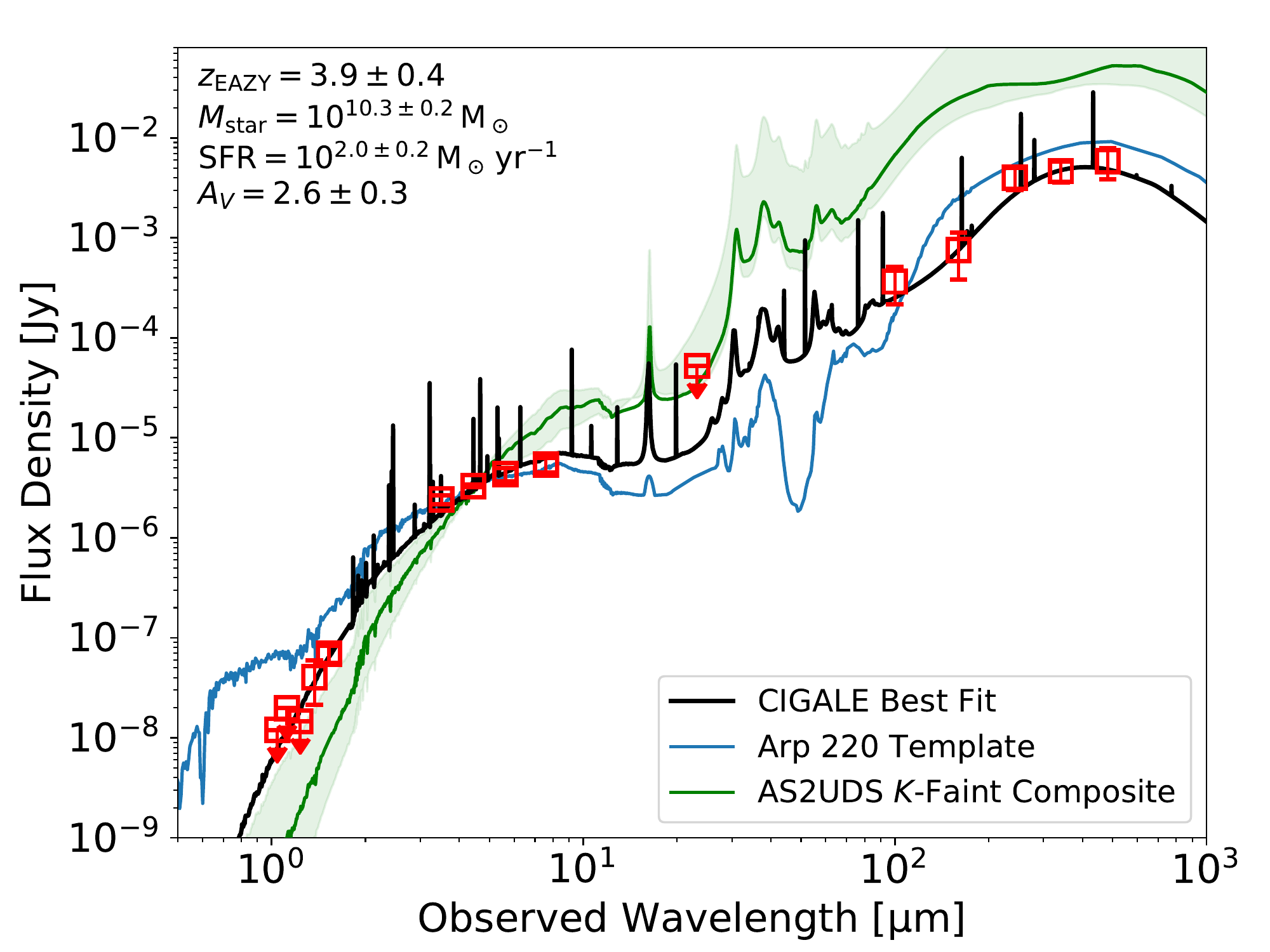}
\caption{\textit{Top-left}: Cutout images of ES-009, the brightest source in our sample, from the near-IR to far-IR. 
In each image, the band name is indicated at the upper-right corner, and the measured brightness is noted at the lower-right corner (in AB magnitude if $\lambda \leq 8$\,\micron, and in flux density if $\lambda > 8$\,\micron).
Numbers in brackets at the lower-left corners are the numbers of images stacked (one for ES-009). 
Scale bars of various lengths are plotted in the left of the cutout images.
\textit{Top-right}: Best-fit SED of ES-009 obtained with \textsc{cigale} (the black solid line).
The photometric data are plotted as open red squares.
For comparison we also plot the SED templates of Arp\,220 (ULIRG; the blue line), mid-IR AGN (\citealt{kirkpatrick15}; the orange line) and AS2UDS $K$-faint SMGs (\citealt{smail20}; the green line with shaded region for the 16--50--84th percentiles of the distribution).
All the templates are redshifted to the same \zphot\ of ES-009 and normalized to the source flux density at 4.5\,\micron, except for the AGN template, which is normalized to the 24\,\micron\ flux density. 
Derived physical properties with \textsc{cigale} ($M_\mathrm{star}$, SFR and $A_V$; lensing magnification corrected) are noted at the upper-left corner.
\textit{Bottom}: Stacked images of all the other 52 sources in the sample (left) and the best-fit SED obtained with \textsc{cigale} (right). 
The layouts are the same as those of the top panels.
}
\label{fig:stack}
\end{figure*}

Among the 53 $H$-faint galaxies, we have identified a few notable sources for which we obtained further measurements. These sources are:

ES-009 --- This source is the brightest (CH2=$20.48\pm0.03$ mag) $H$-faint source found in our sample (Figure~\ref{fig:stack}). 
It is also brighter than any $H$-faint source found in the CANDELS fields \citep[the brightest is at CH2$=$21.96;][]{wang19}, any $K>25.3$ SMG reported in the UDS field \citep[the brightest is at CH2$=$22.24;][]{smail20,dudzevic20}, and any optical/near-IR-dark ALESS SMGs reported in the ECDFS field \citep[the brightest is at CH2$=$22.27;][]{simpson14}.
Stacking the four-band \hst\ data (F105W, F125W, F140W, and F160W) of this source yields no detection.
Furthermore, we detected the IR counterpart of this source in the MIPS 24\,\micron\ and all the five \herschel\ bands (100--500\,\micron; see Figure~\ref{fig:stack}).
According to the mid-IR active galactic nucleus (AGN) template by \citet{kirkpatrick15}, the AGN contribution to its far-IR SED should be less than 10\%.
The far-IR continuum of this source peaks around 350\,\micron\ ($S_{350} = 52\pm 4$\,mJy), suggesting a likely redshift solution of $z\sim 3$.

ES-005, ES-028, ES-029, ES-039 --- Among all the 21 sources that were observed with the \spitzer/MIPS at 24\,\micron\ (Table~\ref{tab:02_phot}), these four sources together with ES-009 are the only cases that are detected in the MIPS 24\,\micron\ band.
The measured MIPS 24\,\micron\ flux densities are 46$\pm$8\,\si{\micro Jy} (ES-005), 263$\pm$11\,\si{\micro Jy} (ES-028), 134$\pm$15\,\si{\micro Jy} (ES-029) and 67$\pm$11\,\si{\micro Jy} (ES-039).
If these sources reside at $z > 3$, then the MIPS detections may imply the presence of hot dust component dominated by AGN \citep[cf.][]{wang16,alcalde19}, and if that is the case, the determination of their photometric redshifts and stellar masses may be inaccurate.

\begin{figure*}[!ht]
\centering
\includegraphics[width=\linewidth]{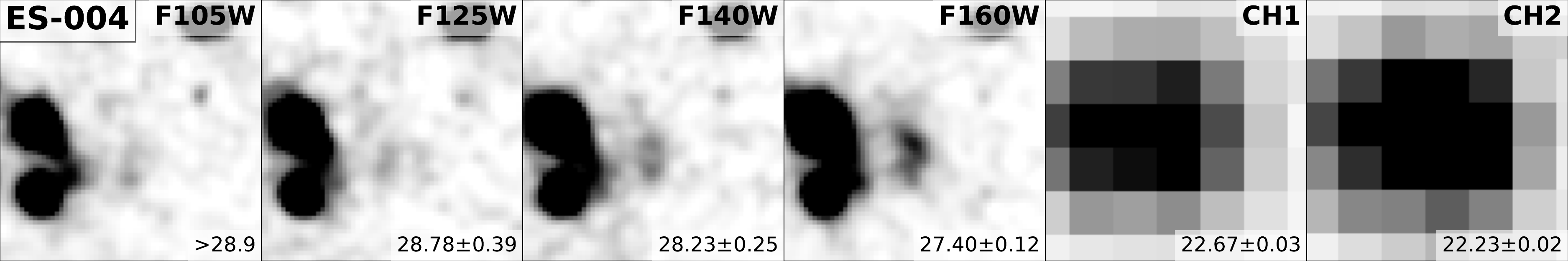}
\includegraphics[width=\linewidth]{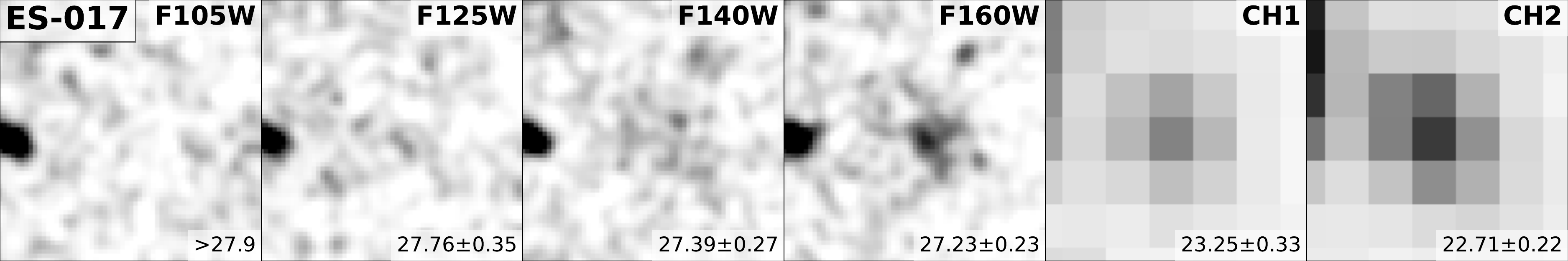}
\caption{Deep cutout images of ES-004 (in A2744) and ES-017 (in MACS0416).
Here the displayed \hst\ images are HFF data \citep{lotz17}, which are deeper than those of BUFFALO and other fields by $\sim 4\times$ (in terms of RMS noise).
\hst\ images are smoothed with a Gaussian kernel (FWHM$=$0\farcs2). 
The band name of each image is indicated at the upper-right corner, and the source magnitude is shown at the lower-right corner.
The magnitude limits are at $3\sigma$.
The size of each cutout image is 3\farcs6$\times$3\farcs6.
}
\label{fig:hff}
\end{figure*}

ES-004, ES-017 --- These two sources are located in the coverage of the original HFF data.
We therefore conducted aperture photometry using the deep HFF data in the WFC3-IR F105W, F125W, F140W and F160W bands with the same method described in Section~\ref{ss:03a_source}.
With the HFF data, whose RMS noise is only $\sim25\%$ of the median RMS noise of our sample, we securely detected ES-004 {and ES-017} at 1.6\,\micron, and both sources are marginally detected at $\lambda_\mathrm{obs} \gtrsim 1$\,\micron\ (Figure~\ref{fig:hff}).
These two sources are not special from the others in terms of intrinsic physical properties.

\subsection{Stacking}
\label{ss:03d_stack}

We conducted stacking analysis to constrain the spectral energy distributions (SEDs) of the $H$-faint sources. 
This is because the studies on the intrinsic properties of individual sources are hampered by the non-detections in the \hst\ and \herschel\ bands.
We split the sources into subsamples by their observed properties.
These groups are 
({\romannumeral 1}) all the 52 sources except for ES-009 which is firmly detected in all the available bands at $\lambda \geq 3.6$\,\micron,
({\romannumeral 2}) four sources (ES-005, ES-028, ES-029 and ES-039) which are detected at 24\,\micron\ (hereafter MIPS-bright sample),
({\romannumeral 3}) 16 sources which are undetected at 24\,\micron\ (hereafter MIPS-faint sample),
({\romannumeral 4}) 13 resolved sources excluding ES-009,
and ({\romannumeral 5}) 39 unresolved sources. 

For sources within each group, we stacked their images in each band with inverse variance weighting of the images.
Similar to \citet{wang19}, we first normalized all the images by the IRAC CH2 flux densities of the sources, and then multiplied the resulting fluxes of stacked source with the median IRAC CH2 flux density.
To mitigate the effect of source confusion in the \herschel\ bands, where the angular resolution is coarse, the images that we stacked are residual images from which nearby sources are removed. 
In these images, we first {identified} \herschel\ sources with the \textsc{daofind} routine \citep{daofind}, and then subtracted all the S/N$>$3 sources that are at least half beam FWHM away from any $H$-faint source of interest using the PSF models.

We note that we used the BUFFALO-only \hst\ data of ES-004 and ES-017 for the stacking of \hst\ images, instead of the deep HFF ones shown in Figure~\ref{fig:hff}. 
This is because with the inverse variance weighting, the weight of deep HFF images are $\sim 16$ times that of the others', and thus these two sources would dominate the signal in the final stack.

We measured the flux densities of the stacked sources with aperture photometry. 
We adopted aperture sizes of $r = 0\farcs6$ in all the \hst\ bands, $1\farcs8$ in the IRAC CH1/2, $2\farcs4$ in the IRAC CH3/4, 7\arcsec\ in the MIPS 24\,\micron, and 5, 10, 18, 27, 36\arcsec\ at 100--500\,\micron. 
Aperture correction factors were calculated based on PSF models for {all \spitzer\ and \herschel\ bands}, and photometric uncertainties were first calculated from the RMS of the stacked background.
{For aperture corrections in the \hst\ bands, we assumed an exponential source profile with $R_\mathrm{e,circ}=0\farcs25$. 
For ES-009 and the stack of the IRAC-resolved sources, we assumed the $R_\mathrm{e,circ}=0\farcs45$ model.
These assumptions were to address the extended source profile and potential smearing effect because of the astrometric uncertainty.
}
If the source is detected at S/N$>$3 in the stacked image, we evaluate the uncertainty with bootstrapping (same method as \citealt{schreiber15} and \citealt{wang19} for the stacking in the \herschel\ bands). 
Therefore, the derived flux density uncertainties incorporate the intrinsic dispersion of sources stacked. 
Table~\ref{tab:03_stack} presents the photometric properties of the stacked sources.

Figure~\ref{fig:stack} shows the stacked images of the 52 sources in all 15 bands. 
The stacked source can be firmly detected in the majority of the bands, except for non-detections in the WFC3-IR/F105W, F110W, F125W and marginal detections in the WFC3-IR/F140W and PACS 100\,\micron\ bands.
The stellar continuum of the stacked source peaks around IRAC/CH4 and the dust continuum peaks around SPIRE 500\,\micron, both favouring a redshift solution around $z \sim 4$.
Although the stacked source is firmly detected at 24\,\micron, the large uncertainty of flux density suggests that the MIPS flux can be contributed and biased by a few MIPS-bright sources.
Further stacking analysis of the 16 MIPS-faint sources suggests a $3\sigma$ upper limit of $<17$\,\si{\micro Jy} at 24\,\micron\ with a median CH2 magnitude of $22.8$. 
Such a flux density can be $>$7 times lower than that of the MIPS-bright sources ($127\pm 12$\,\si{\micro Jy}). 

Because the far-IR flux densities of $H$-faint sources may not well correlate with those measured in the IRAC/CH2 band (e.g., \citealt{wang19}), we also tried to stack images without any normalization. 
The derived photometric results show general consistency with the CH2-normalized ones within a $\sim1\sigma$ confidence interval.
Similar consistency can also be found if we adopt median stacking instead of the inverse variance method.

One well-known caveat for stacking the measurements of a likely mixed population is that the derived physical properties can still deviate from the average properties of individual sources.
Stacking sources in a wide redshift range can lead to an artificial broadening of spectral features.
Additionally, if the $H$-faint sample comprises both dusty star-forming galaxies and old passive systems (i.e., below the $1\sigma$ dispersion of the star-forming main sequence as shown in \citealt{wang19}, and thus likely at slightly lower redshifts), then the stacked SEDs may be biased towards the passive population with less dust obscuration in the near/mid-IR but dominated by the active population in the far-IR.
Further constraints on the far-IR luminosities with ALMA or NOEMA observations will be able to decompose such a mixed population by allowing accurate SFR measurements of individual sources.

\section{Analysis and Discussion} \label{sec:04_ana}

\subsection{Photometric Redshift}
\label{ss:04a_sed}

With the SEDs described in Section~\ref{ss:03c_notable} and \ref{ss:03d_stack} through direct photometry (ES-004, ES-009 and ES-017) and stacking, we run SED modeling software to derive the photometric redshift (photo-z, $z_\mathrm{phot}$) of the $H$-faint sources.
We do not model the SEDs of the majority of individual sources because most of them are only detected in two IRAC bands, and the constraints on the dust continua are poor.
{The lensing magnifications are only corrected for physical quantities after SED fitting, and therefore do not enter the calculation of photometric redshifts.}

We first estimate the \zphot\ using the nine-band near/mid-IR ($\lambda \leq 8$\,\micron) SEDs with \textsc{eazy} \citep{brammer08}.
Similar to \citet{wang19}, we use the full set of spectral template library allowing $A_V$ up to 5. 
The best-fit \zphot\ of the stacked 52 sources (excluding ES-009) derived with \textsc{eazy} is {$3.9\pm0.4$}  (uncertainty denotes the 16-84th percentiles of the likelihood distribution), consistent with the median photometric redshift of $H$-faint sources reported in the CANDELS fields ($z_\mathrm{phot,med}=3.9 \pm 0.1$; \citealt{wang19}), ALESS optically faint SMGs ($3.7\pm0.1$; \citealt{dacunha15}), $K$-faint AS2UDS SMGs ($3.5\pm0.1$ for the subset studied in \citealt{smail20}), serendipitous $H$-faint sources detected with the GOODS-ALMA survey ($\sim3.5$; \citealt{franco18, franco20,zhou20}) and the ALMA large program ALPINE ($\sim 3.7$; \citealt{gruppioni20}), as well as the broad redshift range of $z=3 \sim 5$ suggested by \citet{yamaguchi19} based on the IRAC--ALMA and ALMA--JVLA flux ratios.
This is smaller than the median redshift (4.8) for the sample of \citet{alcalde19}, in which the sources are intrinsically fainter (down to CH2=24.5). 
The derived photo-z's of ES-004 ($4.1_{-1.7}^{+0.6}$), ES-017 ($3.7_{-1.0}^{+1.2}$), the stack of the IRAC-unresolved sources ({$3.9_{-0.8}^{+0.9}$}), IRAC-resolved sources ({$2.6_{-0.4}^{+1.4}$}) and MIPS-faint sources ({$3.6_{-0.3}^{+0.4}$}) are consistent with that of the stack of 52 sources, but MIPS-bright sources are likely at a lower redshift ({$2.6\pm0.4$}).
The best-fit model of ES-009 suggests unreasonable redshift at $z\sim 7$ because it is only detected in the IRAC CH1/2 bands (CH3/4 data do not exist).

We also estimate the \zphot\ and derive the physical properties simultaneously using the full near-to-far-IR SEDs with \textsc{magphys+photo-z} \citep{magphys,battisti19}.
\textsc{magphys} is an energy-balance SED modeling software which assumes a continuous delayed exponential star-formation history (SFH) with random starburst and a two-component dust absorption law \citep[i.e., diffuse ISM and stellar birth cloud;][]{cf00}. 
This is the extension of the software that \citet{dacunha15} used for the analysis of optically faint ALESS SMGs reported in \citet{simpson14}, and \citet{smail20} used it for the analysis of $K$-faint SMGs in UDS field \citep[AS2UDS sample;][]{stach19,dudzevic20}.
Note that we adopt a uniform prior distribution of redshift instead of the default one that peaks at $z\sim 2$ \citep{battisti19}.

The best fit \textsc{magphys} \zphot\ for the stack of the 52-source sample is {$2.7_{-0.4}^{+1.0}$} (uncertainty denotes the 16-84th percentiles of the likelihood distribution).
The large uncertainty on the derived \zphot\ may imply an intrinsically broad redshift distribution of $H$-faint galaxies (likely $z\simeq 2 -5$). 
This is supported by the redshift distribution of individual sources seen in  \citet{wang19} and \citet{gruppioni20}.
We also derive the photometric redshift for ES-009 ({$3.2\pm0.1$}), the stack of the MIPS-bright ($1.8_{-0.2}^{+0.3}$) and MIPS-faint sample ({$3.2\pm0.6$}), respectively. 
The redshifts derived with \textsc{magphys} are generally lower but marginally consistent with those obtained with \textsc{eazy}.
We conclude that the majority of $H$-faint galaxies selected in this work are at similar median redshifts ($z_\mathrm{med} \simeq 3.5-4$) reported in previous studies, although a small fraction of them can be located at a lower redshift of $z\simeq2-3$.

\subsection{Lens Models and Magnification}
\label{ss:04b_lensing}

We utilize published cluster lens models to infer the lensing magnification and thus the intrinsic physical properties.
These are available for 24 out of the 32 clusters, including six HFF clusters, six CLASH clusters and 12 RELICS clusters.
For sources detected in the HFFs, we use the lens models produced by the CATS team \citep{jauzac14,jauzac15b,jauzac16,limousin16,mahler18,lagattuta19} because of a larger field coverage at a high resolution (0\farcs2). 
We also use the large-area low-resolution lens model produced by \citet{merten11} in MACS0416 and MACS1149, where the identified sources are out of the coverage of the CATS products.
For sources detected in the CLASH and RELICS cluster fields, we use all of the available lens models on the MAST high-level science product database, including those produced by CATS \citep{richard14}, Zitrin-dPIEeNFW and Light-Traces-Mass \citep{zitrin13,zitrin15}, and GLAFIC {\citep{oguri10,kawamata16,okabe20}}.

We assume a point-source model for the extraction of magnification factor ($\mu$). 
Based on the photometric redshifts presented in Section~\ref{ss:04a_sed}, we estimate the lensing magnifications for most of the sources at $z = 4$.
The magnifications for the four MIPS-bright sources are estimated assuming $z = 2.6$, and for ES-009 we assume $z=3.2$.
If multiple lens models are available for the same source, we adopt the geometric mean of all magnifications.
As summarized in Table~\ref{tab:02_phot}, lensing magnifications are available for 42 sources in our sample.
In Appendix~\ref{apd:depth}, we show the magnifications versus distances to cluster center of these sources as the top-right panel of Figure~\ref{fig:depth}.
Only two sources are found at $\mu>5$, indicating that lensing-boosted rest-frame UV-optical offset of high-redshift dusty galaxies \citep[e.g.,][]{fujimoto16} is not the cause of non-detection in the F160W band.
We also note that we may preferentially miss strong galaxy-galaxy lensed objects because of a selection bias against sources within 1\arcsec\ of the foreground lens.

The 16--50--84th percentiles of the magnification distribution are 1.3--2.2--3.0.
We uniformly assume $\mu_\mathrm{med}=2.2$ for 11 sources without available lens models.
This is also consistent with the median magnification of sources at comparable distances to the cluster centers ($r=0\farcm9_{-0.4}^{+0.3}$).
For each individual source with available lens models, we estimate a typical magnification uncertainty of $\sigma_\mu / \mu \sim 20$\% based on the standard deviation of magnifications predicted by different models.
The magnification uncertainty introduced by the redshift uncertainty (i.e., $\sigma_{z}/(1+z)\sim20\%$) is found to be a minor effect ($\sigma_\mu / \mu \sim 5$\% at $\mu_\mathrm{med}=2.2$).

With these lens models, four multiply lensed systems can be identified in our sample, including ES-006/07/08 (in MACSJ0032.1+1808), ES-018/19 (in MACSJ0417.5-1154), ES-028/29 (in A697) and ES-045/46 (in MACS2129).
This further reduces the number of independent sources in our sample to 48.

\subsection{Statistics of Physical Properties}
\label{ss:04c_phys}

\begin{figure*}[!ht]
\centering
\includegraphics[width=\linewidth]{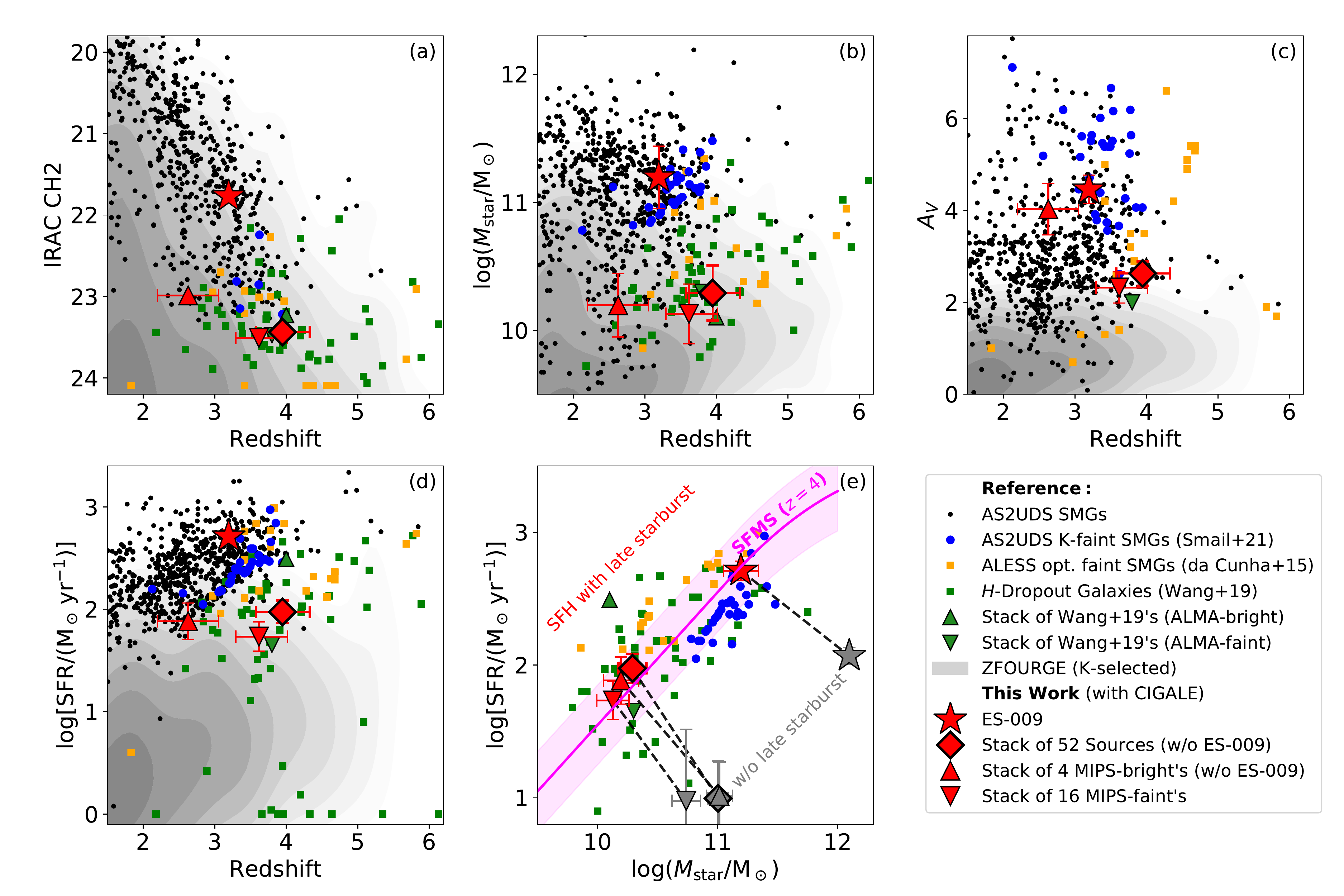}
\vspace{-5mm}
\caption{Physical properties of $H$-faint galaxies (from panel a to d: IRAC CH2 magnitude, stellar mass, rest-frame $V$-band attenuation, SFR versus redshift; panel e: SFR versus stellar mass). Lensing magnification has been corrected.
Sources in our sample are shown as red symbols with black edges in all panels, including ES-009 (star), stack of the other 52 sources (diamond), stack of the MIPS-bright sources (upward triangle) and MIPS-faint sources (downward triangle).
Reference samples include ({\romannumeral 1}) submillimeter-selected AS2UDS SMGs \citep[black dots;][]{dudzevic20}, AS2UDS SMGs at $K>25.3$ \citep[blue circles;][]{smail20}, optically faint ALESS SMGs \citep[orange squares;][]{simpson14,dacunha15}, ({\romannumeral 2}) IRAC-selected $H$-faint galaxies in CANDELS fields \citep[green squares for individual sources, upward/downward triangles for the stacks of ALMA-detected/undetected sources;][]{wang19}, and ({\romannumeral 3}) $K$-band-selected ZFOURGE galaxies \citep[grey contours in the background at arbitrary levels;][]{straatman16}.
In panel (e), we also plot the star-forming main sequence and its $1\sigma$ dispersion at $z=4$ \citep[magenta solid line;][]{schreiber15}. 
The differences between the $M_\mathrm{star}$ and SFR derived with \textsc{cigale} (with a late starburst in the SFH; red-filled symbols) and \textsc{magphys} (without mandatory late starburst; gray-filled symbols) are indicated by the black dashed lines.
}
\label{fig:stats}
\end{figure*}

We find that \textsc{magphys} derives an abnormally low ratio of $\mathrm{SFR}/L_\mathrm{IR}$  ({$10^{-10.6}$\,\si{M_\odot.yr^{-1}.L_\odot^{-1}}}) for all of the $H$-faint source groups. This is only a {third} of that observed for optically faint SMGs at similar redshifts ($10^{-10.1}$\,\si{M_\odot.yr^{-1}.L_\odot^{-1}}; e.g., \citealt{dacunha15,smail20}) and the \citet{kennicutt98} conversion factor assuming a Chabrier IMF, which is likely caused by a large contribution of diffuse dust emission to $L_\mathrm{IR}$ as assumed by \textsc{magphys}. 
To address this issue, we further model the SEDs with the energy-balance code \textsc{cigale} \citep{cigale09,cigale19} assuming a continuous delayed SFH with a recent starburst in the past 20\,Myr.
This is to interpret the dust and reddened stellar continuum simultaneously with a dusty starburst and an old population. 
We also assume a modified \citet{calzetti00} extinction law (allowing a steeper power-law slope index up to $0.4$; cf.\ \citealt{salim20}), and nebular emission is included in the modeling.
We use \textsc{eazy}-derived $z_\mathrm{phot}$ for all the cases except for ES-009, where the \textsc{magphys}-derived one is adopted.
We run Monte-Carlo sampling of photometric redshifts based on their probability distributions and then feed into \textsc{cigale} fitting routine, and thus the uncertainties of derived physical properties incorporate the uncertainties of redshifts.
{We note that \textsc{cigale} is not designed as a photo-z code. However, based on the $\chi_\nu^2$ of best-fit SED models of the same source at different redshifts, we find that $\exp[-\chi_\nu^2(z)]$ can be well modeled with a Gaussian function peaking at the same $z_\mathrm{phot}$ as that derived with \textsc{eazy} for most of the cases except for ES-009, whose probability function peaks around $z=3.5$. 
}

With \textsc{cigale}, we derive a median stellar mass of {$10^{10.3\pm 0.3}$\,\msun}, SFR of {$100_{-40}^{+60}$\,\smpy}\ and rest-frame $V$-band attenuation of {$A_V = 2.6 \pm 0.3$} for the stacked 52 sources in our sample after correcting for the median magnification.
Similar $M_\mathrm{star}$ and SFR are derived for the stacked MIPS-bright and MIPS-faint sources.
However, we find that the MIPS-bright sources show a larger $A_V$ ({$4.0\pm0.6$}) over the MIPS-faint sample ({$2.3\pm0.3$}), highlighting that the $H_{160}-\mathrm{CH2}$ selection technique is subject to a degeneracy of $z_\mathrm{phot}-A_V$ \citep[also pointed out in][]{caputi12,wang16}.
The brightest source ES-009 is found to host a stellar mass of {$10^{11.2\pm0.2}$}\,\msun\ and SFR of {$500_{-180}^{+290}$}\,\smpy\ with an $A_V$ of $4.5\pm 0.3$.
A summary of physical properties can be found in Table~\ref{tab:03_stack}.

Figure~\ref{fig:stats} shows the physical properties of optical/near-IR-dark galaxies presented in several works.
We find that the assumption of SFH holds the key to the interpretation of \mstar\ and SFR in $H$-faint galaxies. 
With the addition of the late starburst that can contribute $22\pm9$\% of the stellar mass in the \textsc{cigale} modeling, the derived SFR is {7$\pm$1} times higher than that of \textsc{magphys}-derived one but \mstar\ is {6$\pm$1} times lower (see panel e of Figure~\ref{fig:stats}).
We favor the \textsc{cigale} results because of a more commonly adopted $\mathrm{SFR}/L_\mathrm{IR}$ ratio and a lower \mstar\ that is compatible with the observed galaxy stellar mass function (SMF) at $z\sim4$ \citep[e.g.,][]{muzzin13,song16}. 
Additionally, the measured S\'ersic indices of resolved $H$-faint sources ($n\sim0.8$) do not resemble those of quenching galaxies as indicated by \textsc{magphys} (i.e., $\gtrsim$10 times below the star-forming main sequence as shown in Figure~\ref{fig:stats}, panel e; \citealt{schreiber15}).

We note that the goodness of fit with the two softwares are comparable ($\chi_{\nu}^2 \sim 1$), and assuming the same extinction law \citep{cf00} does not mitigate the conspicuous difference in either SFR or \mstar.
{Similarly, the SFR and \mstar\ modeled with \textsc{magphys} (\citealt{dacunha15}; optimzed for high-redshift applications) using the same $z_\mathrm{phot}$ derived with \textsc{eazy} are consistent with those derived with \textsc{magphys+photo-z} instead of \textsc{cigale}.
}
Recent studies on massive galaxies at $z<1$ also suggest that parametric SFH modeling may underestimate the uncertainties of SFR, \mstar\ when compared to those with non-parametric/stochastic SFH assumptions \citep[e.g.,][]{leja19,iyer20}, and it is possible that the true SFR and \mstar\ of $H$-faint galaxies are between the \textsc{cigale} and \textsc{magphys} fitting results.
This underscores the difficulty of physical property characterization with limited information in the optical/near-IR bands.

We also assess the impact of the late burst age on the SED modeling results.
Assuming a 100-Myr late burst (i.e., $\sim$10\% of the maximum age of these galaxies, which is consistent with their number density among the broader population; see Section~\ref{ss:04f_sfrd}), we find that with a $\sim$40\% increase of mean $\chi^2_\nu$ to $\sim 1.5$, the derived SFR (\mstar) of all groups will slightly decrease (increase), but still within the $1\sigma$ uncertainties of those quantities derived with a 20-Myr late burst. 
Similarly, adopting lower \zphot\ derived with \textsc{magphys} will also lead to a small decline of estimated SFR and \mstar, which are also within the uncertainties summarized in Table~\ref{tab:03_stack}.
This further suggests that the uncertainties of derived physical properties of $H$-faint galaxies are dominated by systematic factors (e.g., assumption of SFH).

The stellar mass and $A_V$ of $H$-faint sources in our sample are consistent with those reported by \citet{wang19} using the same software.
The SFR of the stacked sources in our sample is between those of ALMA-detected and undetected $H$-faint galaxies in \citet{wang19}, suggesting that the sample consists of galaxies across the wide dispersion of the so-called star-forming main sequence at $z\sim4$ ($\sigma\sim0.3$\,dex; \citealt{schreiber15}).
Compared with submillimeter-selected optical/near-IR-dark galaxies in \citet{dacunha15}, \citet{dudzevic20} and \citet{smail20}, sources in our sample are generally lower in SFR and $A_V$ because 850\,\micron\ selection favors dust-rich galaxies with high SFR.
A lower \mstar\ is derived for our sample despite comparable 4.5\,\micron\ flux densities, and this can be interpreted by the difference in dust extinction and SFH assumption.

{Assuming $z_\mathrm{phot}=3.9$ for the stack of 52 sources, we derive a dust temperature of $40\pm6$\,K from a modified blackbody fit to the Herschel/SPIRE flux densities (dust emissivity $\beta=1.5$). 
We note that the Herschel data alone cannot provide a tight constraint on dust temperature because of rising SPIRE flux densities as a function of wavelength and the artificial broadening due to stacking.
However, the derived $T_\mathrm{dust}$ is generally consistent with the $T_\mathrm{dust}$ of $L_\mathrm{IR}=10^{12}$\,\lsun\ galaxies at $z\simeq3-4$ as reported by \citet[][$\sim$38\,K]{schreiber18b}, as well as the $T_\mathrm{dust}$ of stacked $H$-faint galaxies in \citet[][$\sim$37\,K]{wang19} assuming the same $\beta$.
}

\subsection{Resolved stellar continuum}
\label{ss:04d_mor}

\begin{figure}[!t]
\centering
\includegraphics[width=\linewidth]{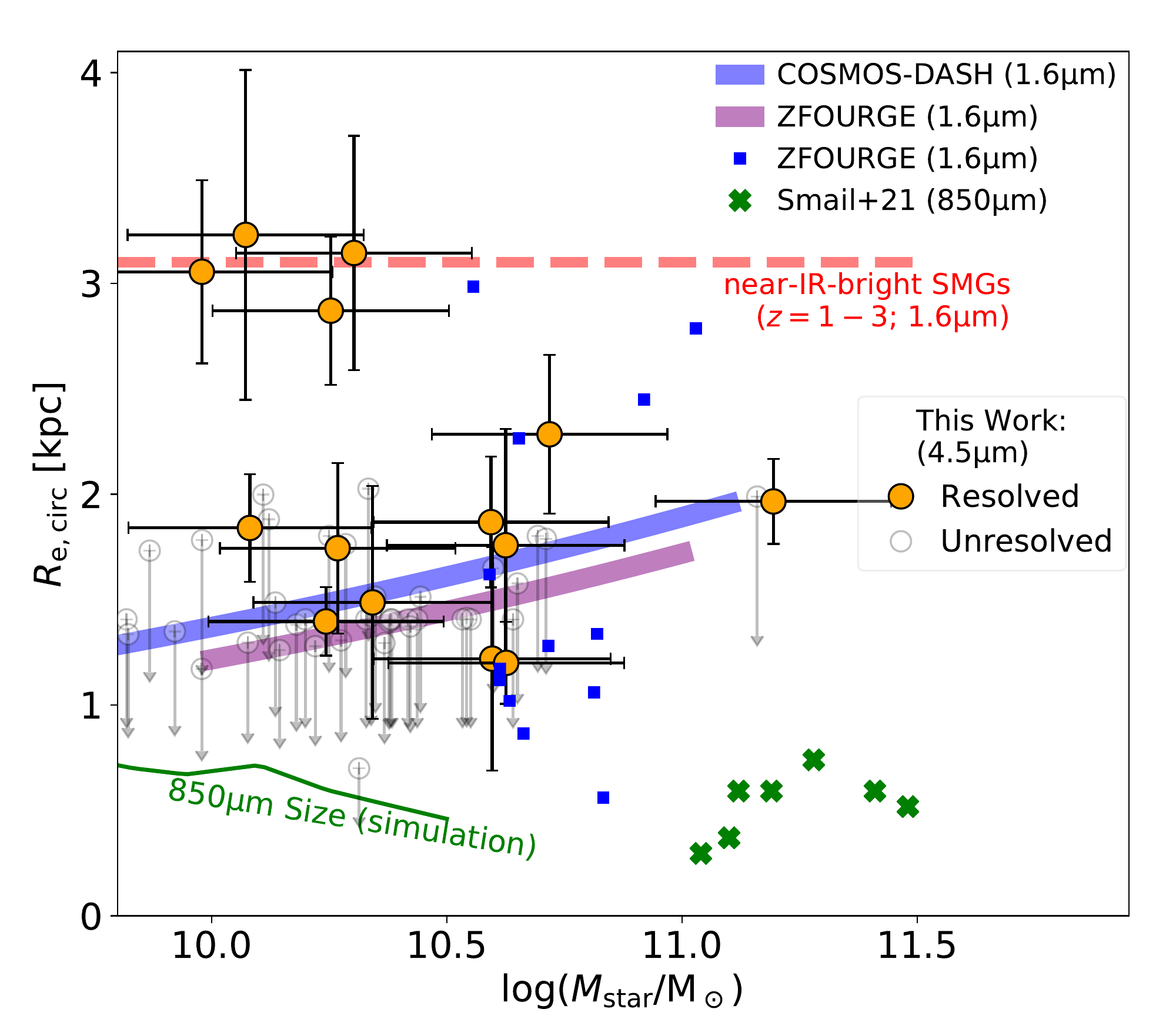}
\vspace{-5mm}
\caption{Circularized effective radii versus stellar masses. 
The orange filled circles denote resolved $H$-faint galaxies in this work, and the transparent open circles represent the upper limits for unresolved sources, assuming $R_\mathrm{e,circ} < 0\farcs3\,\mu^{-0.5}$.
The reference samples that we plot include ({\romannumeral 1})  star-forming ZFOURGE galaxies at $z\sim3.8$ (the blue squares: \citealt{straatman15}; the purple line, \citealt{allen17}; 1.6\,\micron\ sizes), 
({\romannumeral 2}) star-forming galaxies at $z=4$ inferred from COSMOS-DASH \citep[the blue line;][scaled from $z=2.75$]{mowla19},
and ({\romannumeral 3}) near-IR-bright ALESS SMGs at $z\simeq1-3$ \citep[the red dashed line;][1.6\,\micron\ sizes]{chen15}.
We also compare the 850\,\micron\ dust continuum sizes and stellar masses of $K$-faint SMGs (the green crosses; observed by \citealt{smail20}) and star-forming galaxies (the green solid line; simulated by \citealt{popping21}) at similar redshift.
}
\label{fig:r_mass}
\end{figure}

We find that the IRAC-resolved $H$-faint galaxies are generally at a higher lensing magnification ($\mu_\mathrm{med} = 2.8\pm0.4$) than the unresolved sample ($\mu_\mathrm{med} = 2.0\pm0.2$), which is likely a selection effect.
After correcting the lensing magnification, we find no clear difference of stellar mass or SFR between the two samples.

Figure~\ref{fig:r_mass} shows the circularized effective radii versus stellar mass of $H$-faint galaxies in our sample.
For those unresolved sources, we assume upper limits of their angular size at $R_\mathrm{e,circ} < 0\farcs3\,\mu^{-0.5}$ (i.e., FWHM less than one pixel).
Stellar mass of each source is computed using the IRAC/CH2 flux density assuming the same intrinsic SED of the stacked source.
We adopt a conversion factor of 0.144\,\si{\arcsec.kpc^{-1}} from the angular to physical scale at $z=4$, which is insensitive to the variation of redshift.
For the 14 resolved sources, we measured a median $R_\mathrm{e,circ}$ of $1.9\pm0.2$\,kpc, and the typical upper limit of 39 unresolved sources is $R_\mathrm{e,circ} < 1.5$\,kpc.
Therefore, we conclude that the typical size of (less obscured) stellar continua of $H$-faint galaxies should be less than 1.5\,kpc, slightly lower than the $R_\mathrm{e}=2$\,kpc size obtained on the stacked F160W images of galaxies at $H_{160} - {m}_{4.5} > 2.25$ in the sample of \citet[note that F160W non-detection was not required for this sample]{wang16}. 
This is much smaller than the F160W size of ALESS SMGs at $z=1\sim3$ ($R_\mathrm{e,circ}\sim 3.1$\,kpc assuming median $b/a=0.5$; \citealt{chen15}) in which no dependence on redshift was identified, suggesting no significant extended stellar disk component ($R_\mathrm{e}\gtrsim 3$\,kpc) in the majority of $H$-faint galaxies unlike those in near-IR-bright SMGs.

We note that although IRAC/CH2 samples the rest-frame $z$-band stellar continuum emissions for $H$-faint galaxies at $z\sim4$, it is still possible that a more compact and obscured stellar component could remain undetected at the dust continuum centroid \citep[e.g.,][]{simpson17as2uds,lang19,sun21}.
Therefore, the true stellar mass profile can be more compact than the 4.5-\micron\ light profile.
The likely existence of such a fully dust-obscured stellar component is also consistent with previous findings that the $A_V$ of SMGs can be underestimated through energy-balance SED modeling \citep[e.g.,][]{casey14b}.

\citet{gullberg19} measured a median dust continuum size of $R_\mathrm{e,circ}=0.6\pm0.1$\,kpc for seven $K$-faint SMGs at 870\,\micron\ (\citealt{smail20}; note that extended halos/disks might not be bright enough to be detected individually with these works). 
This is found to be more compact than those of near-IR-bright SMGs at similar redshifts with comparable $M_\mathrm{star}$ and $M_\mathrm{dust}$ but lower $A_V$. 
As also suggested in \citet{smail20}, we argue that the compactness of both dust and stellar continua ($R_\mathrm{e}\sim 1$\,kpc) are likely the cause of high dust attenuation inferred from the SEDs of $H$-faint galaxies.
With no significant evolution in the dust continuum size versus redshift \citep[e.g.,][]{gullberg19,tadaki20} and $\sim (1+z)^{-1}$ evolution in the stellar continuum size \citep[e.g.,][]{mosleh12,vdw14,mowla19}, SMGs at a higher redshift ($z\sim 4$) are more likely to host compact dust and stellar continua of similar size than those at lower redshifts ($z=1\sim2$ where the difference in $R_\mathrm{e,star}$ and $R_\mathrm{e,dust}$ can be around 2\,kpc; e.g., \citealt{hodge16}, \citealt{lang19}, \citealt{franco20}, \citealt{sun21}), producing highly obscured stellar continua in the optical/near-IR bands.
Most recently, \citet{popping21} also reported a comparable size of stellar and dust continua for star-forming galaxies at $z\sim4$ using TNG50 simulation.
Further ALMA continuum observations on these cluster-lensed $H$-faint galaxies will test this interpretation directly.

We do not observe any clear correlation between the effective radii and stellar masses of resolved $H$-faint sources. 
The four sources at $R_\mathrm{e,circ} > 2.5$\,kpc are less massive (mean {$M_\mathrm{star} = 10^{10.2 \pm 0.1}$\,\msun}) than the compact but resolved sample ({$10^{10.5\pm 0.2}$\,\msun}).
Sources with relatively large $R_\mathrm{e,circ}$ may represent galaxies in the final coalescence phase of merger (e.g., see the saddle-like shape in the contours of ES-016; Figure~\ref{fig:all_sources}).
Such an interpretation can be tested with future high-resolution ALMA or \textit{JWST} observations.

We compared the $R_\mathrm{e,circ}-M_\mathrm{star}$ distribution with massive star-forming galaxies ($>10^{10}$\,\msun) at $z\sim 4$ reported in the literature, including the ZFOURGE sample \citep{straatman15,allen17} and the COSMOS-DASH sample 
(\citealt{mowla19}; note that we extrapolate the best-fit relation at $z=2.75$ to $z = 4$ by a scale factor of $(1+z)^{-1}$ and assume a typical axis ratio of $b/a = 0.5$). 
We find that the effective radii of $H$-faint galaxies are in good agreement with those of $H$-selected star-forming galaxies.
This suggests that $H$-faint galaxies represent the massive and dusty tail of the distribution of the wider galaxy population at $z\sim 4$ instead of a distinct class.

\subsection{Number Density of $H$-faint sources}
\label{ss:04e_dens}

\begin{figure}[!t]
\centering
\includegraphics[width=\linewidth]{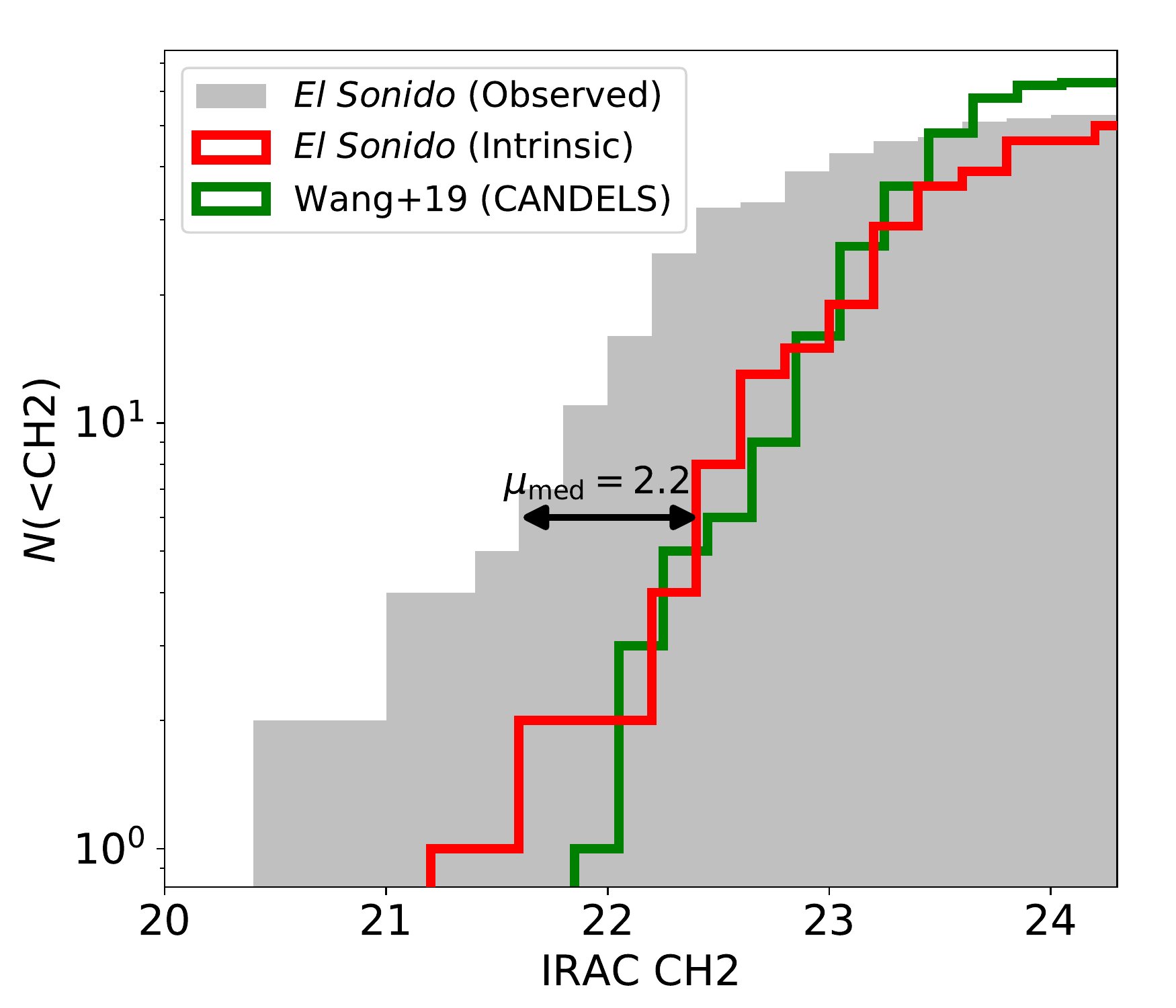}
\vspace{-5mm}
\caption{Cumulative number counts of $H$-faint sources presented in this work and \citet[][green solid line]{wang19} as a function of IRAC CH2 magnitude.
The grey filled histogram denotes the cumulative distribution of directly observed CH2 magnitude (i.e., without lensing correction), and the red solid line denotes the distribution of magnification-corrected CH2 magnitude. 
Although we discover $\sim$10 times more bright $H$-faint sources at CH2\,$<22$ than \citet{wang19}, the source count as a function of lensing-corrected CH2 brightness is consistent with that derived in the CANDELS fields.
}
\label{fig:hist_ch2}
\end{figure}

\begin{figure*}[!t]
\centering
\includegraphics[width=0.49\linewidth]{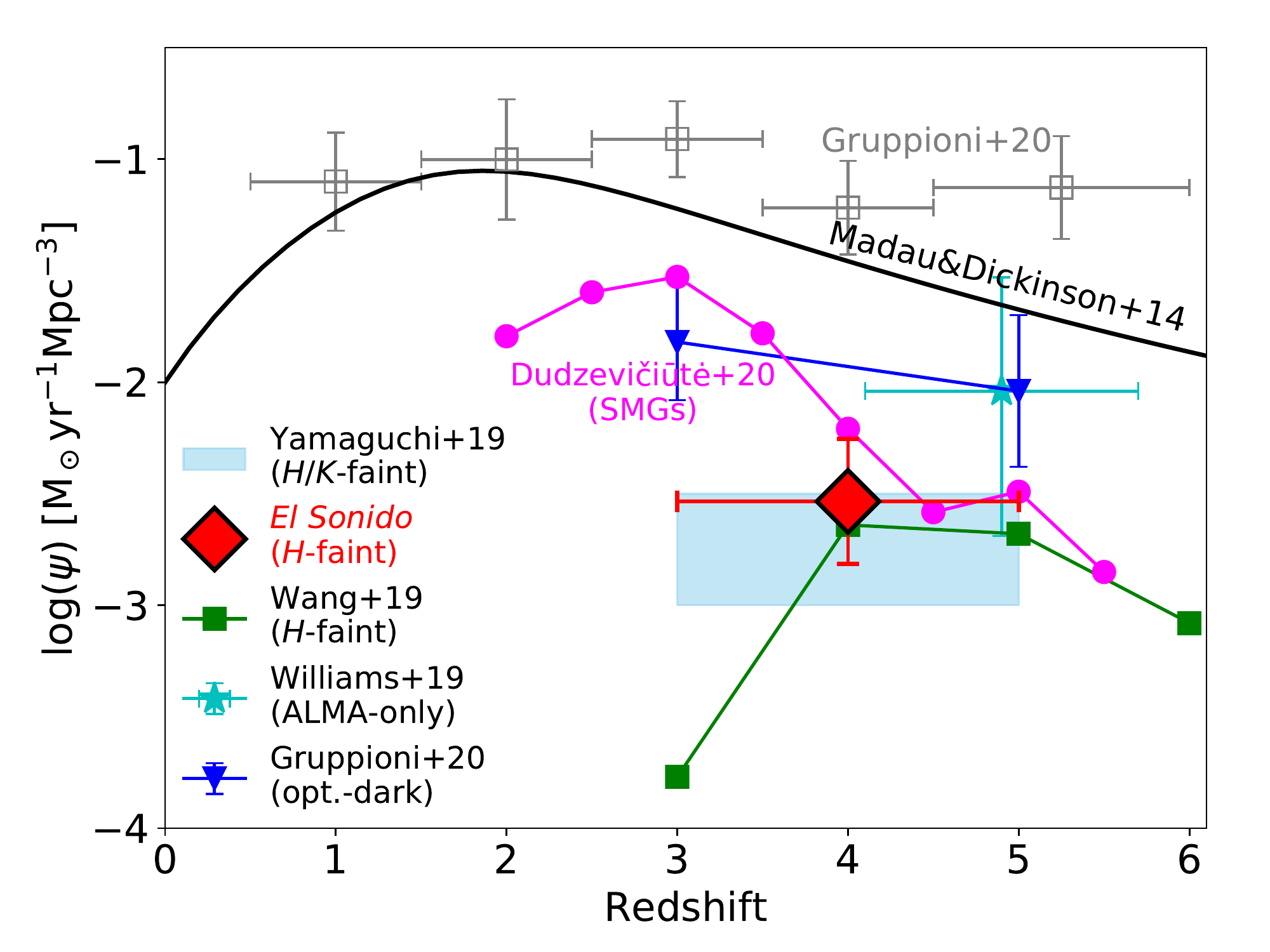}
\includegraphics[width=0.49\linewidth]{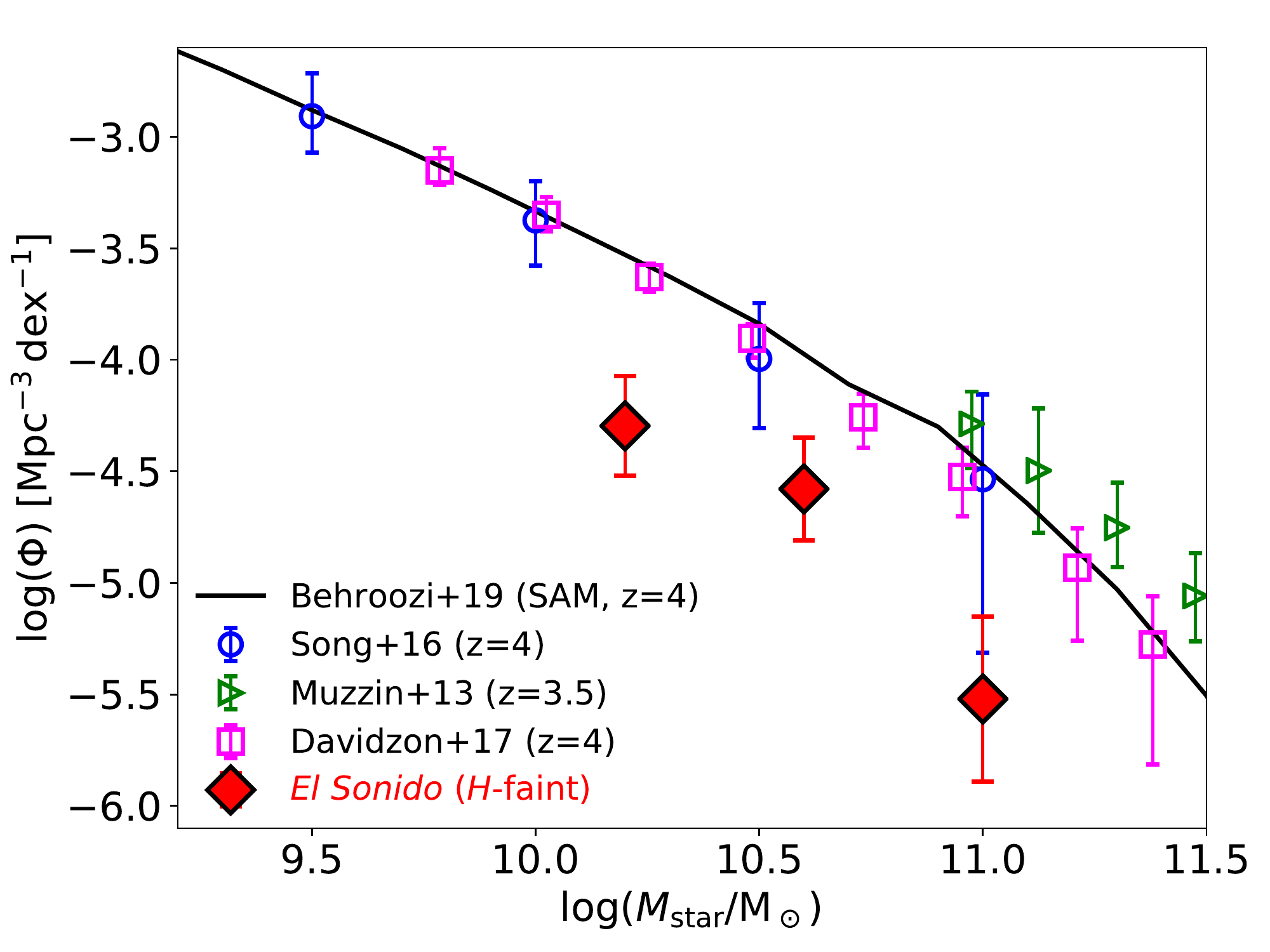}
\vspace{-3mm}
\caption{
\textit{Left}: CSFRD contribution of $H$-faint galaxies presented in this work (the red diamond).
CSFRD contribution of $H$-faint galaxies found by \citet{yamaguchi19} and \citet{wang19}, optical/near-IR-dark galaxies found by \citet{williams19} and \citet{gruppioni20}, as well as AS2UDS SMGs ($S_{870}>1$\,mJy; \citealt{dudzevic20}) are also shown for comparison.
The ALMA-based CSFRD evolution measured by \citet{gruppioni20} is shown as grey open squares.
The analytic model by \citet{md14} is shown as black solid line.
\textit{Right}: The contribution of $H$-faint galaxies to the galaxy SMF at $z\sim4$. 
Unobscured galaxies selected by \hst\ (\citealt{song16}; the blue open circles), VISTA/$K_S$ (\citealt{muzzin13}; the green open triangles) and $K + \mathrm{CH1}$ (\citealt{davidzon17}; the magenta open squares) are shown for comparison. 
Black solid line denotes the SMF derived with a semi-analytic model \citep{behroozi19}.
}
\label{fig:csfr}
\end{figure*}

Figure~\ref{fig:hist_ch2} shows the cumulative number count of $H$-faint galaxies presented in this work (lensed sample) and \citet[][unlensed sample]{wang19} as a function of IRAC CH2 magnitude.
Compared with this blind survey in the CANDELS fields {(the effective survey area is $\sim300$\,arcmin$^2$)}, we have discovered $\sim 10$ times more bright $H$-faint galaxies at CH2$<$22 owing to the lensing magnification. 
After correcting for magnification of each source, the two number count functions are similar, especially in the range of $22 <\mathrm{CH2} < 23.5$.
At the faint end (CH2\,$>23.5$), the source count derived in our sample is lower. 
This is because the detection of intrinsically faint sources requires a larger lensing magnification for a smaller survey area (see the lower-right panel of Figure~\ref{fig:depth} in Appendix~\ref{apd:depth}).

We then estimate the surface number density of $H$-faint galaxies.
The apparent total survey area of the 101 cluster fields in the F160W band is summed to be 648\,arcmin$^2$. 
With the estimated completeness ratio ($C\sim76$\%) and the cumulative effective survey area as a function of intrinsic CH2 magnitude threshold (i.e., $A_\mathrm{eff}(\mathrm{CH2})$; on the source plane) as detailed in Appendix~\ref{apd:area}, we derive a maximum intrinsic survey area of $\sim$221\,arcmin$^2$. 
This can be further reduced to $\sim$120\,arcmin$^2$ at CH2$=$24 because a larger magnification is required to detect fainter sources, limiting the effective survey area.
Based on the differential number count and effective survey area, both as functions of intrinsic CH2 magnitude, we derive a surface number density of $H$-faint sources (intrinsic $\mathrm{CH2}\leq 24$ and $H_{160}\gtrsim 27.2$) of $(7.9\pm2.2)\times 10^2$\,deg$^{-2}$.
Note that each multiply imaged system has been treated as one source in this calculation.

Assuming an ALMA detection rate ($S_{870}>0.6$\,mJy) of $62\pm 6$\% as reported in \citet{wang19}, we would expect a surface density of ALMA-bright $H$-faint galaxies as $(4.9\pm1.4)\times 10^2$\,deg$^{-2}$.
This is consistent with the measured density of ALMA-detected $H$-faint galaxies ($\sim530$\,deg$^{-2}$, \citealt{wang19}) and $K > 25.3$, $S_{870} \geq 1$\,mJy SMGs ($450^{+750}_{-300}$\,deg$^{-2}$, \citealt{smail20}).

\subsection{Contribution to the CSFRD and SMF}
\label{ss:04f_sfrd}

If $80\pm9$\% of the $H$-faint galaxies (i.e., the fraction of MIPS-faint sources) are in the redshift range of $z \simeq 3 - 5$, we can then estimate their contribution to the cosmic SFR density (CSFRD) as $\psi = 10^{-2.5\pm 0.3}$\,\si{M_\odot.yr^{-1}.Mpc^{-3}}, corresponding to $8_{-4}^{+8}$\% of the CSFRD at this epoch \citep{md14}.
{Survey completeness has been corrected for this CSFRD calculation.}
As shown in the left panel of Figure~\ref{fig:csfr}, such a contribution is consistent with that of $H$-faint galaxies reported previously \citep{yamaguchi19,wang19}, and also comparable to that of AS2UDS SMGs at $z\sim4$ ($S_{870}>1$\,mJy, i.e., $L_\mathrm{IR}>10^{12}$\,\lsun) reported in \citet{dudzevic20}.

We further notice that the contribution to the CSFRD from $H$-faint galaxies in this work is $\sim 3$ times lower than the values suggested by \citet{williams19} and \citet{gruppioni20} through serendipitous ALMA detections in relatively smaller survey areas (one source in 8\,arcmin$^2$ and seven sources in 25 arcmin$^2$, respectively).
An average near-IR magnitude of 23.7$\pm$0.1 can be measured from the stacked UltraVISTA $Y J  H K_S$ data for the sample of \citet{gruppioni20}, suggesting the potential detectability of these sources with deeper F160W data (e.g., at a depth comparable to those of this work, \citealt{yamaguchi19} and \citealt{wang19}), 
and thus the estimate of CSFRD may be consistent.
Furthermore, as argued by \citet{zavala21}, possible clustering of serendipitous sources around main ALPINE targets can also be a concern in \citet{gruppioni20} sample.

We also evaluate the contribution to the SMF at $z\sim 4$ from the $H$-faint galaxies in the right panel of Figure~\ref{fig:csfr}.
In the three stellar mass bins from {$10^{10}$ to $10^{11.2}$\,\msun}, $H$-faint galaxies contribute to {$16_{-7}^{+13}$\%} of the volume densities {($10^{-4.3\pm0.2}$, $10^{-4.6\pm0.2}$ and $10^{-5.5\pm0.4}$\,\si{Mpc^{-3}.dex^{-1}})} that are derived based on unobscured near-IR-selected galaxies at comparable redshift \citep{muzzin13,song16,davidzon17}, consistent with the fraction reported by \citet{alcalde19}.
We note that \citet{caputi15} and \citet{alcalde19} suggested a higher contribution of $K$/$H$-faint IRAC sources to the most massive ($M_\mathrm{star} > 10^{11}$\,\msun) galaxy population at $z>4$.
Although {only two sources in our sample (ES-009 and ES-050) are found} in this parameter space because of the limited survey area, different selection criteria and assumption of SFH, {we may tentatively estimate a fraction of $H$-faint galaxies at $M_\mathrm{star}>10^{11}$\,\msun\ as $15_{-8}^{+21}$\%, consistent} with the value reported by \citet[27$\pm$17\% at $4<z<5$]{alcalde19}.

% \section{Discussion} \label{sec:05_discuss}
% \input{05_dis}

\section{Summary} \label{sec:05_sum}

We have conducted a \spitzer/IRAC survey of $H$-faint ($H_{160} \gtrsim 26.4$, $<$\,5$\sigma$) galaxies in 101 lensing cluster fields over a total survey area of 648\,arcmin$^2$ (effectively $\sim$221\,arcmin$^2$ on the source plane), including all of the 68 best studied massive galaxy clusters observed with the \hst\ Large/Treasury Programs: HFF (BUFFALO), CLASH and RELICS.
Using imaging data obtained with \hst/WFC3-IR, \spitzer/IRAC and MIPS, \herschel/PACS and SPIRE, we have carried out photometric measurements, surface brightness profile modeling, stacking analysis and near-to-far-IR (1.05--500\,\micron) SED modeling of the selected $H$-faint galaxies.
The main results of this study are the following:

\begin{enumerate}

\vspace{-2mm}
\item We have detected 53 (48 independent) $H$-faint galaxies in 32 cluster fields.
With a median IRAC/CH2 magnitude of $22.46\pm0.11$, all of these sources show a red near-IR color of $H_{160} - \mathrm{CH2} > 2.5$ (the median is above 3.9) {assuming a point-source spatial profile}. 
The median IRAC CH1--CH2 color is $0.49 \pm 0.03$, suggesting a reddened stellar continuum.

\vspace{-2mm}
\item We were able to resolve 14 out of the 53 sources in the IRAC CH2 band thanks to the gravitational lensing.
The mean circularized effective radius of the resolved sources is $0\farcs46\pm0\farcs04$, corresponding to a physical scale of $1.9\pm0.2$\,kpc at $z=4$ after lensing magnification correction.
Considering the unresolved sources, the median stellar continuum size of $H$-faint galaxies is very compact ($R_\mathrm{e,circ}<1.5$\,kpc).

\vspace{-2mm}
\item 
We estimate the photometric redshift of $H$-faint galaxies with the stacked near-IR SEDs.
The majority ($80\pm9$\%) of $H$-faint galaxies are faint in the MIPS 24\,\micron\ band ($<$17\,\si{\micro Jy}), and the derived {$z_\mathrm{phot}=3.6_{-0.3}^{+0.4}$} is consistent with those of optical/near-IR-dark galaxies reported previously.
A small fraction of sources in this study ($20\pm9$\%) are bright at 24\,\micron\ ($127\pm12$\,\si{\micro Jy}), and they are likely located at a lower redshift ({$z_\mathrm{phot}=2.6\pm0.4$}).

\vspace{-2mm}
\item
After correcting for the lensing magnification, we show that the $H$-faint galaxies in this study are massive ({median $M_\mathrm{star} = 10^{10.3\pm 0.3}$\,\msun}), strongly star-forming ({$\mathrm{SFR} = 100_{-40}^{+60}$\,\smpy}) and highly dust-obscured ({$A_V = 2.6 \pm 0.3$}).
The assumption of star formation history (i.e., with or without a late starburst) is critical for the characterization of SFR and stellar mass, which can result in a factor of {$\sim 7$ times} difference.

\vspace{-2mm}
\item
The most remarkable source in our sample, ES-009 in the field of A2813, is the brightest $H$-faint galaxy at 4.5\,\micron\ known so far (CH2=$20.48\pm0.03$).
This galaxy is detected in all eight available bands at $\lambda \geq 3.6$\,\micron\ up to 500\,\micron, and likely hosts a stellar mass of {$10^{11.2\pm0.2}$\,\msun}, SFR of {$500_{-180}^{+290}$\,\smpy}\ at $z_\mathrm{phot}=3.2\pm0.1$.

\vspace{-2mm}
\item
We conclude that the highly obscured stellar continuum of $H$-faint galaxies at $z\sim 4$ is likely caused by the compactness of both the stellar and dust components ($R_\mathrm{e,circ}\sim 1$\,kpc; see \citealt{smail20}).
The stellar continuum sizes of $H$-faint galaxies are in general agreement with those of massive unobscured galaxies at similar redshifts, suggesting that $H$-faint galaxies represent the massive and dusty tail of the distribution of the wider galaxy population at $z\sim4$.

\vspace{-2mm}
\item
We derive a sky surface density of 
$(7.9\pm2.2)\times 10^2$\,\si{deg^{-2}} for \cite{wang19}-like $H$-faint galaxies in lensing cluster fields.  This number is consistent with the reported density of $H$-faint galaxies measured in blank fields \citep[CANDELS;][]{wang19} and $K$-faint SMGs (\citealt{smail20}) if $\sim$40\% of $H$-faint galaxies are less active and faint at 870\,\micron\ ($S_{870}<0.6$\,mJy).
We further conclude that $H$-faint galaxies contribute to $8_{-4}^{+8}$\% of the cosmic SFR density at $z=3\sim 5$, and account for {$16_{-7}^{+13}\%$} of the galaxies in the stellar mass range of {$10^{10}-10^{11.2}$\,\msun}\ at this epoch.

\end{enumerate}
\vspace{-2mm}

This extensive lensing survey of optical and near-infrared dark objects (\elsonido) yields a bright sample of $H$-faint galaxies, owing to cluster lensing magnification, with rich ancillary data from the optical to far-IR wavelengths.
Future observations with ALMA and \textit{JWST} will provide key insights of the dust and (less obscured) stellar continua of these dusty and massive galaxies at $z\sim4$, which are typically missed in \hst-based surveys or certain ``stellar-mass''-selected galaxy samples, improving our understanding of their physical properties (i.e., dust mass, dust temperature, dust/stellar continuum size) and formation history.

% \vspace{-0.5cm}

\acknowledgments

{We thank the anonymous referee for helpful comments.
FS and EE acknowledge funding from JWST/NIRCam contract to the University of Arizona, NAS5-02105.}
PGPG acknowledges support from Spanish Government grant PGC2018-093499-B-I00.
IRS acknowledges support from STFC (ST/T000244/1).
KIC acknowledges funding from the European Research Council through the award of the Consolidator Grant ID 681627-BUILDUP.
GEM acknowledges the Villum Fonden research grant 37440, ``The Hidden Cosmos" and the Cosmic Dawn Center of Excellence funded by the Danish National Research Foundation under then grant No.\ 140. 
MJ is supported by the United Kingdom Research and Innovation (UKRI) Future Leaders Fellowship ``Using Cosmic Beasts to uncover the Nature of Dark Matter'' (grant number MR/S017216/1).

% Spitzer SHA
This work is based on observations made with the Spitzer Space Telescope, which was operated by the Jet Propulsion Laboratory, California Institute of Technology under a contract with NASA.

% HST archive
This work is based on observations made with the NASA/ESA Hubble Space Telescope, obtained from the data archive at the Space Telescope Science Institute.

% HFF lens models
This work utilizes gravitational lensing models produced by PIs Brada\v{c}, Natarajan \& Kneib (CATS), Merten \& Zitrin, Sharon, Williams, Keeton, Bernstein and Diego, and the GLAFIC group. This lens modeling was partially funded by the HST Frontier Fields program conducted by STScI. STScI is operated by the Association of Universities for Research in Astronomy, Inc.\ under NASA contract NAS 5-26555. The lens models were obtained from the Mikulski Archive for Space Telescopes (MAST).

% Herschel
This work is based on observations made with Herschel.
Herschel is an ESA space observatory with science instruments provided by European-led Principal Investigator consortia and with important participation from NASA.

%% To help institutions obtain information on the effectiveness of their 
%% telescopes the AAS Journals has created a group of keywords for telescope 
%% facilities.
%
%% Following the acknowledgments section, use the following syntax and the
%% \facility{} or \facilities{} macros to list the keywords of facilities used 
%% in the research for the paper.  Each keyword is check against the master 
%% list during copy editing.  Individual instruments can be provided in 
%% parentheses, after the keyword, but they are not verified.

% \vspace{5mm}
\facilities{HST(WFC3), Spitzer(IRAC and MIPS), Herschel(PACS and SPIRE)}

%% Similar to \facility{}, there is the optional \software command to allow 
%% authors a place to specify which programs were used during the creation of 
%% the manuscript. Authors should list each code and include either a
%% citation or url to the code inside ()s when available.

\software{astropy \citep{astropy}, CIGALE \citep{cigale09,cigale19}, EAZY \citep{brammer08}, GALFIT \citep{galfit}, MAGPHYS \citep{magphys,battisti19}, Photutils \citep{photutils}, SExtractor \citep{sex}
}

%% Appendix material should be preceded with a single \appendix command.
%% There should be a \section command for each appendix. Mark appendix
%% subsections with the same markup you use in the main body of the paper.

%% Each Appendix (indicated with \section) will be lettered A, B, C, etc.
%% The equation counter will reset when it encounters the \appendix
%% command and will number appendix equations (A1), (A2), etc. The
%% Figure and Table counter will not reset.

% \clearpage

\appendix

% \vspace{-5mm}

% \restartappendixnumbering

\begin{figure*}[!t]
\centering
\begin{minipage}{0.54\linewidth}
\includegraphics[width=\linewidth]{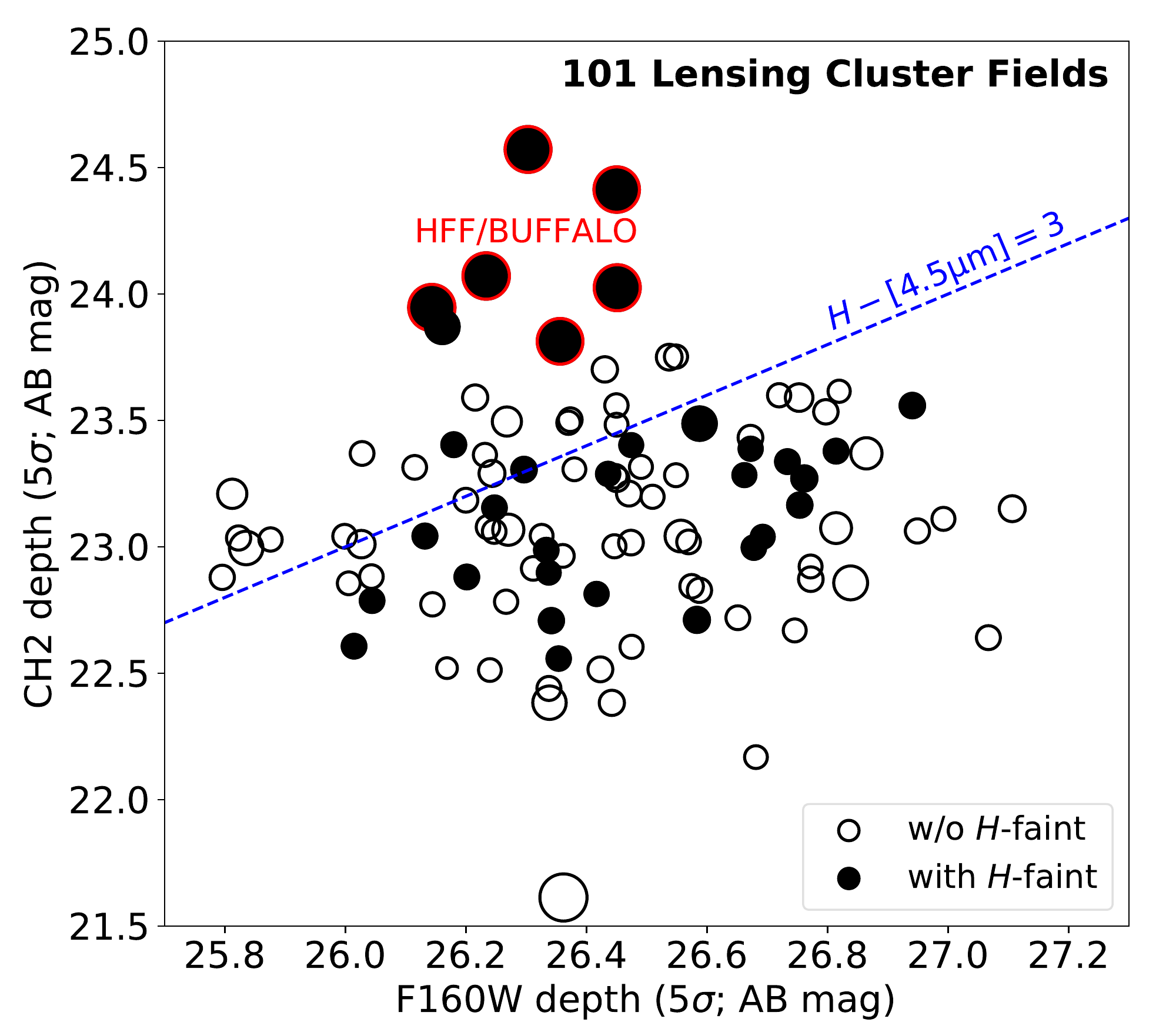}
\end{minipage}\hfill
\begin{minipage}{0.45\linewidth}
\includegraphics[width=\linewidth]{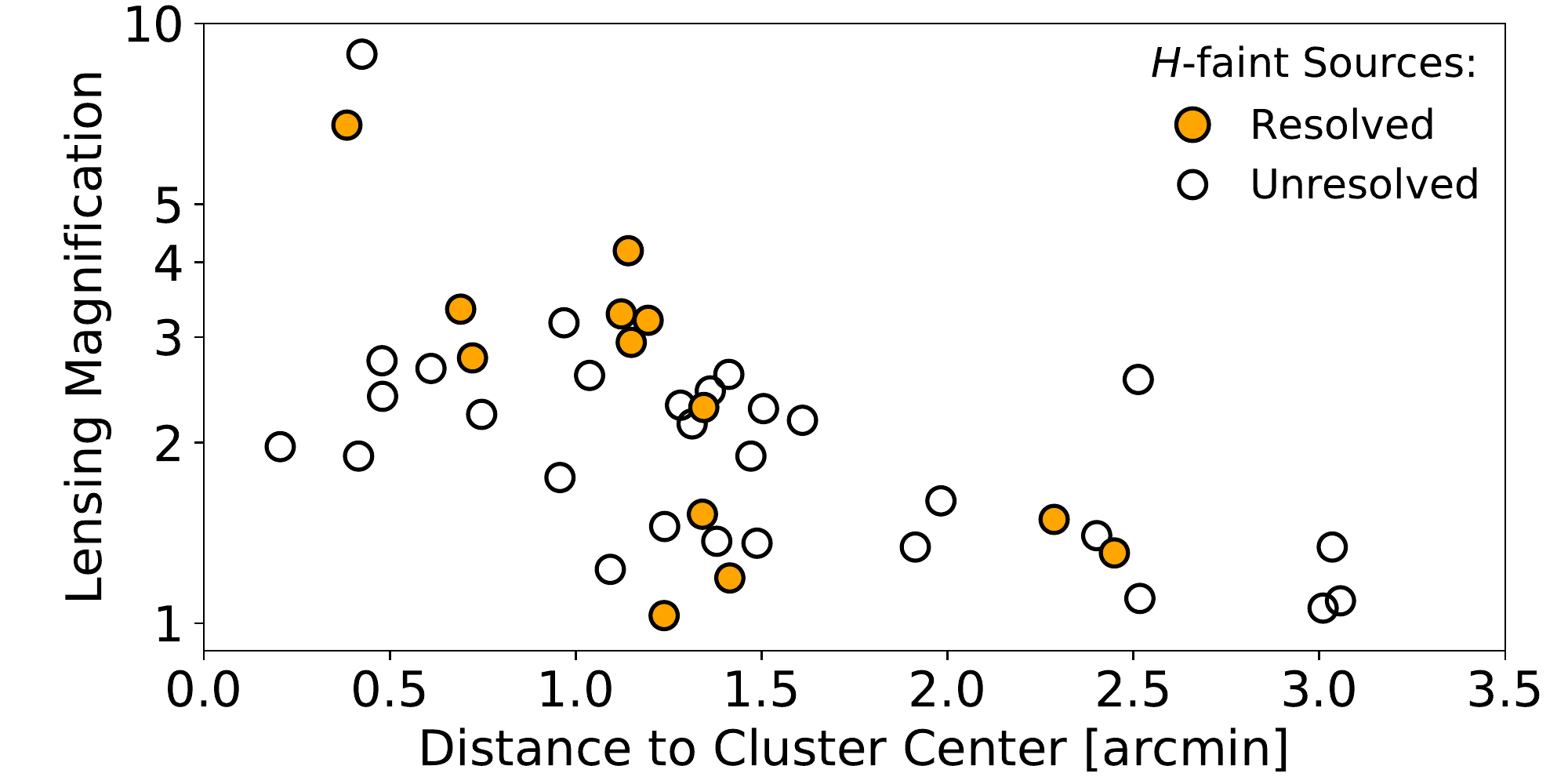}\hfill
\includegraphics[width=\linewidth]{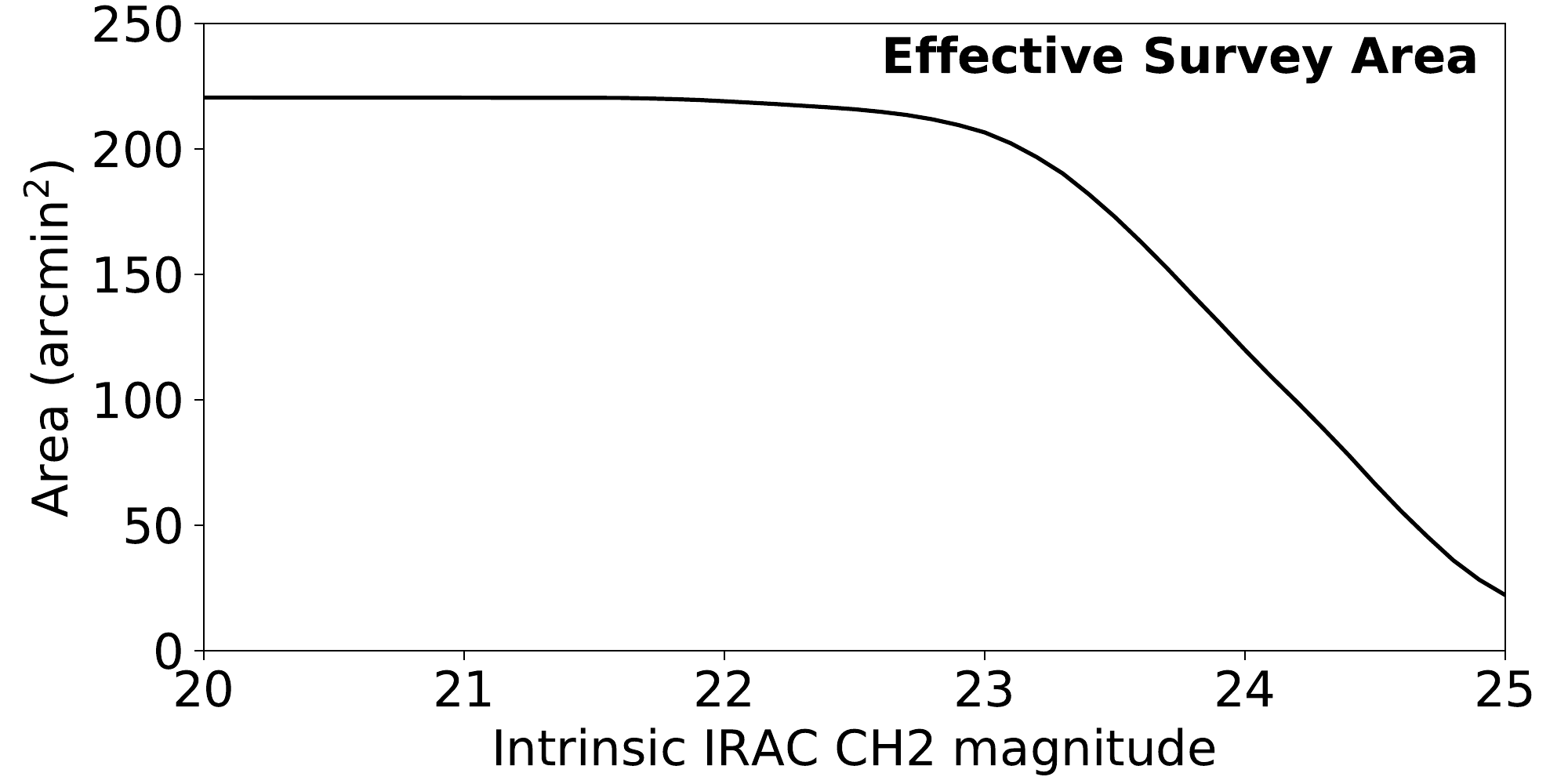}
\end{minipage}
\vspace{-3mm}
\caption{\textit{Left}: Depths of imaging data in the IRAC/CH2 and WFC3-IR/F160W band. 
32 cluster fields in which we detected $H$-faint galaxies are shown as black filled circles, and the remaining 69 fields are shown as empty ones. 
Larger circle sizes indicate larger apparent survey areas.
Six HFF/BUFFALO clusters are marked as circles with red edges.
$H$-faint galaxies discovered in clusters below the blue dashed line are automatically at $H_{160} - \mathrm{CH2} > 3$.
\textit{Top-right}: Lensing magnifications of $H$-faint galaxies versus their distances to the corresponding cluster centers.
Orange filled circles denote resolved sources, and open circles denote unresolved ones. 
\textit{Bottom-right}: Cumulative effective survey area in the source plane as a function of intrinsic IRAC/CH2 magnitude threshold. 
The effective survey area is 120\,arcmin$^2$ for sources at an intrinsic depth of CH2=24.
}
\label{fig:depth}
\end{figure*}

\section{Depths of the 101 Lensing Cluster Fields}
\label{apd:depth}

In each lensing cluster field, the $5\sigma$ depths of the WFC3-IR/F160W and IRAC/CH2 data in their overlapped area are measured using randomly distributed apertures. 
The aperture sizes {and aperture-correction factors} are identical to those that we adopted to obtain photometric measurements of $H$-faint sources.
Table~\ref{tab:01_cluster} summarizes the depths of data in all 101 cluster fields.
The 16--50--84 percentiles of the $5\sigma$ depths are 26.2--26.4--26.7 in the F160W band and 22.8--23.1--23.5 in the CH2 band. 
The left panel of Figure~\ref{fig:depth} shows the distribution of depths in these two bands.
Excluding the six HFF/BUFFALO clusters, where the apparent survey area per cluster field is about four times the typical ones (i.e., with only one WFC3-IR pointing), we find no clear difference in the data depth among the cluster fields with and without $H$-faint galaxies through a Kolmogorov–Smirnov (K--S) test. 

We also evaluate the influence of F160W or CH2 source densities on $H$-faint galaxy selection.
We have run \textsc{SExtractor} in these fields and found a median surface density of $\Sigma_\mathrm{1.6} = 0.057$\,arcsec$^{-2}$ and $\Sigma_\mathrm{4.5} = 0.014$\,arcsec$^{-2}$ for F160W and CH2 sources at S/N$>5$, respectively.
A K--S test suggests no difference in source densities among cluster fields with and without $H$-faint galaxies.
The cumulative number of $H$-faint galaxies versus the angular distance to the cluster center can be well characterized with a simple quadratic function ($N\propto r^2$) at $r<1\farcm4$, i.e., within the FoV of one WFC3-IR pointing.
This further suggests that the increasing source density towards the cluster center does not have a significant influence on the $H$-faint source selection.

\section{Completeness and Intrinsic Survey Area}
\label{apd:area}

We first estimate the completeness of our $H$-faint galaxy selection.
Because we only surveyed $H$-faint sources that are $>$1\arcsec\ away from \hst\ sources and distinguishable from other IRAC sources (i.e., $>$1\farcs8 away according to the Rayleigh criterion), we correct the survey area in each field by a factor of $\exp[- (\Sigma_{1.6}-\Sigma_{4.5}) \pi r_{1.6}^2 - \Sigma_{4.5} \pi r_{4.5}^2]$ where $r_{1.6} = 1$\arcsec\ and $r_{4.5}=1\farcs8$.
Based on the elliptical parameters of sources derived with \textsc{SExtractor}, we find that bright and extended ($R_\mathrm{e,maj} \gtrsim 1$\arcsec) galaxies in the fields can only lead to a minor reduction of the effective survey area by $\sim 1\%$.
We then derive an effective survey area of 490\,arcmin$^2$ in the image plane.
This implies that $\sim 24$\% of the $H$-faint galaxies will be missed in our survey due to the existence of nearby field sources, for example, the triply imaged $H$-faint ALMA-bright galaxy at $z=4.3$ behind the cluster 0102-4915 \citep{caputi21}.

{
We also note that $H$-faint sources are often found with less-obscured companions detectable with \hst\ \citep[e.g.,][]{simpson17,schreiber18,caputi21}.
Adopting the angular cross-correlation function between $H$-faint and rest-frame-UV-selected galaxies at similar photometric redshifts  reported in \citet{wang19} and the surface density of $H<27.3$ galaxies at $z\simeq3-5$ within the GOODS field \citep{barro19}, the surface density of UV-bright companion galaxies within 1\arcsec\ from $H$-faint sources is around 0.004\,arcsec$^{-2}$, i.e., one order of magnitude lower than random field sources. 
Therefore, the additional clustering effect of $H$-faint and companion galaxies will not be a concern for our completeness calculation.
}

Based on the lens models described in Section~\ref{ss:04b_lensing}, we derive the cumulative effective survey area in the source plane as a function of magnification threshold (i.e., $A_\mathrm{eff}(\mu)$) in each cluster field.
For cluster fields without available lens models, we used the median of $A_\mathrm{eff}(\mu)$ curves for approximation (scaled to the survey area of these fields).
These functions are further converted to $A_\mathrm{eff}(\mathrm{CH2})$, i.e., functions of intrinsic CH2 magnitude threshold, according to the CH2 depths of corresponding cluster fields.
The cumulative effective survey area, after lensing correction, as a function of intrinsic CH2 magnitude threshold in the 101 cluster fields is shown in the lower-right panel of Figure~\ref{fig:depth}.
With a maximum at $\sim$221\,arcmin$^2$, the intrinsic survey area is reduced to $\sim$120\,arcmin$^2$ at an intrinsic depth of CH2=24.

Based on the synthesized $A_\mathrm{eff}(\mu)$ curve and the 53-source sample size, we anticipate to detect 3.4 sources in the area with a lensing magnification of $\mu > 5$, and 0.5 source with $\mu > 10$.
Both expectations are consistent with the observed numbers of strongly lensed $H$-faint galaxies (two at $\mu>10$ and zero at $\mu>10$).
Therefore, we further conclude that the selection bias against strongly lensed (and therefore extended) sources should be negligible.

% \begin{longrotatetable}
\startlongtable 

\begin{deluxetable*}{@{\extracolsep{2pt}}rrrrrrrrrrr} %[p!]
\tablecaption{\elsonido\ Cluster Sample
\label{tab:01_cluster}}
% \tablenum{01}
\tablewidth{0pt}
\tabletypesize{\scriptsize}
\tablehead{
\colhead{\#} & \colhead{Cluster Name} & \multicolumn{2}{c}{Coordinates} & \multicolumn{4}{c}{\hst/WFC3-IR F160W} & \multicolumn{2}{c}{\spitzer/IRAC CH1\&2} \\
\cline{3-4} \cline{5-8} \cline{9-10}  %\vspace{-2mm}
\colhead{} & \colhead{} & \colhead{RA} & \colhead{Dec} 
& \colhead{TP\tablenotemark{a}} & \colhead{$t_\mathrm{obs}$\tablenotemark{b}} 
& \colhead{Depth\tablenotemark{c}} & \colhead{Program\tablenotemark{d}} &
\colhead{Depth\tablenotemark{e}}  & \colhead{Program\tablenotemark{f}} \\
\colhead{}  & \colhead{} 
& \colhead{(deg)} & \colhead{(deg)} & \colhead{} & \colhead{(h)} & \colhead{(mag)} & \colhead{} & \colhead{(mag)} & \colhead{} 
}
% \decimalcolnumbers
\startdata
  1 &         MACS1149 & 177.388 &  22.391 &     HFF{/BUFFALO} & 40.93{/1.79} & 26.5 & 1,2,3,4,5,6,7,8,9 & 24.4 & 1,2,3  \\
  2 &            A370 &  39.939 &  --1.549 &     HFF{/BUFFALO} & 24.48{/1.79} & 26.1 &      3,8,10,11,12 & 23.9 & 1,4,5,6  \\
  3 &            A2744 &   3.583 & --30.387 &     HFF{/BUFFALO} & 22.35{/1.79} & 26.5 &           8,13,14 & 24.0 & 7,8  \\
  4 &         MACS0416 &  64.018 & --24.088 &     HFF{/BUFFALO} & 22.34{/1.79} & 26.3 &        8,13,15,16 & 24.6 & 9,10  \\
  5 &         MACS0717 & 109.427 &  37.738 &     HFF{/BUFFALO} & 21.90{/1.79} & 26.4 &           8,17,18 & 23.8 & 1,2,11,12  \\
  6 &           AS1063 & 342.204 & --44.536 &     HFF{/BUFFALO} & 21.81{/1.79} & 26.2 &           8,19,20 & 24.1 & 1,7,13  \\
  7 &           BULLET & 104.630 & --55.947 & \nodata &  4.69 & 26.6 &             10,21 & 23.0 & 1,2,14,15  \\
  8 &        0205--5829 &  31.420 & --58.488 & \nodata &  3.30 & 26.8 &          22,23,24 & 22.9 & 16,17  \\
  9 &            A1763 & 203.800 &  41.000 &  RELICS &  2.98 & 26.0 &                25 & 23.0 & 18,19,20,21  \\
 10 &         MACS1423 & 215.951 &  24.074 &   CLASH &  2.65 & 27.1 &             13,26 & 23.2 & 1,2,22  \\
 11 &          RCS2327 & 351.870 &  --2.083 & \nodata &  2.57 & 26.8 &                27 & 23.6 & 1,2  \\
 12 &          RXJ1347 & 206.891 & --11.805 &   CLASH &  2.40 & 26.5 &          10,13,28 & 23.8 & 1,2,7  \\
 13 &   CLJ0152.7--1357 &  28.167 & --13.955 &  RELICS &  2.24 & 26.9 &                25 & 23.6 & 18,19,23,24,25,26,27,28  \\
 14 &        0615--5746 &  93.953 & --57.789 &  RELICS &  2.15 & 26.8 &             25,29 & 22.9 & 18,19,29,30  \\
 15 &        2106--5844 & 316.537 & --58.735 & \nodata &  2.12 & 26.4 &             24,30 & 21.6 & 31  \\
 16 &           MS2137 & 325.051 & --23.639 &   CLASH &  2.03 & 26.8 &             31,32 & 23.3 & 1,7,16  \\
 17 &          RBS1748 & 322.414 &   0.077 &   CLASH &  1.73 & 26.4 &                33 & 22.4 & 1,32,33,34  \\
 18 &        2040--4451 & 310.274 & --44.837 & \nodata &  1.68 & 27.1 &             22,23 & 22.6 & 30,35  \\
 19 &            A383 &  42.020 &  --3.525 &   CLASH &  1.65 & 26.3 &             34,35 & 23.1 & 1,7,16  \\
 20 &        0546--5345 &  86.642 & --53.772 & \nodata &  1.56 & 26.8 &                24 & 23.1 & 31,36  \\
 21 &         MACS1720 & 260.039 &  35.583 &   CLASH &  1.47 & 26.3 &             35,36 & 23.5 & 34,37,38  \\
 22 &         MACS0647 & 101.929 &  70.283 &   CLASH &  1.45 & 26.6 &                37 & 23.0 & 1,39  \\
 23 &            A521 &  73.535 & --10.209 & \nodata &  1.45 & 26.7 &                38 & 23.0 & 1  \\
 24 & MACSJ0940.9+0744 & 145.202 &   7.728 & \nodata &  1.45 & 26.6 &                39 & 22.7 & 40  \\
 25 &          CLJ1226 & 186.734 &  33.550 &   CLASH &  1.43 & 26.9 &                40 & 23.1 & 16,23,26,41  \\
 26 &            A2261 & 260.586 &  32.108 &   CLASH &  1.40 & 26.5 &                41 & 23.0 & 7,32,37  \\
 27 &         MACS0329 &  52.402 &  --2.222 &   CLASH &  1.40 & 26.6 &                42 & 22.8 & 9  \\
 28 & MACSJ1157.3+3336 & 179.296 &  33.599 &   CLASH &  1.40 & 26.2 &                43 & 23.3 & 16,32,40,42  \\
 29 &         MACS2129 & 322.366 &  --7.696 &   CLASH &  1.40 & 26.7 &                44 & 23.3 & 1,2,43  \\
 30 &         MACS1311 & 197.782 &  --3.166 &   CLASH &  1.40 & 26.8 &                45 & 23.5 & 9,44  \\
 31 &            A209 &  22.973 & --13.603 &   CLASH &  1.40 & 26.2 &                46 & 23.6 & 7,9  \\
 32 &         MACS1206 & 181.583 &  --8.785 &   CLASH &  1.40 & 26.3 &                47 & 23.3 & 9  \\
 33 &            A611 & 120.233 &  36.058 &   CLASH &  1.40 & 26.4 &                48 & 23.3 & 1,45  \\
 34 &         MACS0744 & 116.209 &  39.463 &   CLASH &  1.40 & 26.8 &                49 & 23.4 & 1,2  \\
 35 &          RXJ1532 & 233.258 &  30.343 &   CLASH &  1.40 & 26.4 &                50 & 23.7 & 16,33,37,46  \\
 36 &            A1689 & 197.856 &  --1.303 & \nodata &  1.40 & 25.8 &                51 & 23.2 & 7,16,32,47,48  \\
 37 &         MACS1115 & 168.943 &   1.486 &   CLASH &  1.37 & 26.7 &                52 & 23.4 & 9,38  \\
 38 &         MACS1931 & 292.979 & --26.552 &   CLASH &  1.34 & 26.4 &                53 & 22.6 & 9,49  \\
 39 &         MACS0429 &  67.380 &  --2.907 &   CLASH &  1.34 & 26.8 &                54 & 23.2 & 9  \\
 40 &            A2218 & 248.968 &  66.214 & \nodata &  1.34 & 26.5 &             10,55 & 23.2 & 7,47  \\
 41 & MACSJ2214.9--1359 & 333.755 & --13.971 & \nodata &  1.33 & 26.7 &                56 & 23.6 & 1,2  \\
 42 & MACSJ0723.3--7327 & 110.883 & --73.432 &  RELICS &  1.26 & 26.7 &                25 & 23.0 & 18,19,25,34  \\
 43 &            A1758 & 203.173 &  50.574 &  RELICS &  1.13 & 26.2 &                25 & 23.9 & 1,7  \\
 44 &        2344--4243 & 356.191 & --42.693 & \nodata &  1.12 & 26.7 &                57 & 22.7 & 30,46,50  \\
 45 & MACSJ0159.0--3412 &  29.765 & --34.244 & \nodata &  0.83 & 26.4 &                58 & 22.5 & 40  \\
 46 &            49187 & 231.121 &   9.960 & \nodata &  0.80 & 27.0 &                59 & 23.1 & 16,51,52,53  \\
 47 &            A1703 & 198.724 &  51.812 & \nodata &  0.78 & 26.4 &                51 & 23.5 & 54  \\
 48 &          RCS0224 &  36.142 &  --0.043 & \nodata &  0.73 & 26.8 &                60 & 22.9 & 1,55,56,57  \\
 49 &            A68 &   9.288 &   9.151 & \nodata &  0.67 & 26.2 &                10 & 23.4 & 1  \\
 50 & MACSJ0257.6--2209 &  44.402 & --22.172 & \nodata &  0.67 & 26.4 &                61 & 23.0 & 16,45  \\
 51 &           AS1077 & 344.713 & --34.801 & \nodata &  0.67 & 26.5 &                10 & 23.3 & 1,7,48  \\
 52 &            A773 & 139.474 &  51.731 & \nodata &  0.67 & 26.0 &                10 & 22.9 & 1,7,32,45  \\
 53 &            A1835 & 210.267 &   2.816 & \nodata &  0.67 & 26.4 &                10 & 23.5 & 1,4  \\
 54 &           MS0451 &  73.564 &  --3.058 & \nodata &  0.67 & 26.8 &                10 & 23.6 & 1,2,7,43  \\
 55 & MACSJ2201.9--5956 & 330.486 & --59.977 & \nodata &  0.61 & 25.8 &                62 & 22.9 & 40  \\
 56 &           MS1358 & 209.894 &  62.487 & \nodata &  0.61 & 26.5 &                10 & 23.8 & 1,7,54  \\
 57 &             A665 & 127.713 &  65.862 &  RELICS &  0.61 & 25.8 &                25 & 23.0 & 16,18,19  \\
 58 & MACSJ0035.4--2015 &   8.850 & --20.289 &  RELICS &  0.59 & 26.4 &                25 & 23.3 & 18,19  \\
 59 &  PLCKG308.3--20.2 & 229.608 & --81.503 &  RELICS &  0.59 & 26.0 &                25 & 22.9 & 16,18,19,25,34,58  \\
 60 &  PLCKG004.5--19.5 & 289.271 & --33.522 &  RELICS &  0.57 & 26.2 &                25 & 22.5 & 18,19,20  \\
 61 &  PSZ2G138.6--10.8 &  36.779 &  49.008 &  RELICS &  0.57 & 26.7 &                25 & 22.2 & 18,19,34  \\
 62 &  RXCJ0142.9+4438 &  25.721 &  44.639 &  RELICS &  0.57 & 26.1 &                25 & 22.8 & 18,19  \\
 63 & MACSJ0159.8--0849 &  29.945 &  --8.858 &  RELICS &  0.56 & 26.4 &                25 & 23.5 & 16,18,19  \\
 64 & MACSJ0245.5--5302 &  41.367 & --53.065 &  RELICS &  0.56 & 26.1 &                25 & 23.3 & 18,19,36,59  \\
 65 &        0102--4915 &  15.703 & --49.283 &  RELICS &  0.56 & 26.9 &                25 & 23.4 & 18,19,25,36  \\
 66 & MACSJ0417.5--1154 &  64.370 & --11.926 &  RELICS &  0.56 & 26.0 &                25 & 22.8 & 18,19,45  \\
 67 &         MACS0018 &   4.696 &  16.415 &  RELICS &  0.56 & 26.7 &                25 & 22.7 & 7,18,19,60  \\
 68 &    MS1008.1--1224 & 152.600 & --12.644 &  RELICS &  0.56 & 26.2 &                25 & 23.1 & 18,19,25,48  \\
 69 &            A697 & 130.732 &  36.375 &  RELICS &  0.56 & 26.1 &                25 & 23.0 & 1,7,16,61,62  \\
 70 & MACSJ0911.2+1746 & 137.830 &  17.789 &  RELICS &  0.56 & 26.7 &                25 & 23.4 & 1,34,59  \\
 71 &      RXSJ0603+42 &  90.838 &  42.232 &  RELICS &  0.56 & 26.3 &                25 & 22.4 & 18,19,25,32  \\
 72 & MACSJ1615.7--0608 & 243.921 &  --6.165 &  RELICS &  0.56 & 25.8 &                25 & 23.0 & 18,19,32,48  \\
 73 & MACSJ0553.4--3342 &  88.380 & --33.684 &  RELICS &  0.55 & 26.5 &                25 & 23.4 & 18,19,45,59  \\
 74 & MACSJ1514.9--1523 & 228.718 & --15.413 &  RELICS &  0.55 & 26.2 &                25 & 23.4 & 16,18,19  \\
 75 &         WHLJ0137 &  24.354 &  --8.457 &  RELICS &  0.55 & 26.5 &                25 & 23.3 & 16,18,19,49  \\
 76 & MACSJ0032.1+1808 &   8.026 &  18.109 &  RELICS &  0.55 & 26.4 &                25 & 22.8 & 18,19,45  \\
 77 &  PSZ2G209.8+10.2 & 110.596 &   7.408 &  RELICS &  0.55 & 26.2 &                25 & 22.5 & 18,19,49  \\
 78 &            A2813 &  10.859 & --20.611 &  RELICS &  0.55 & 26.2 &                25 & 23.2 & 1  \\
 79 &  PLCKG287.0+32.9 & 177.704 & --28.085 &  RELICS &  0.55 & 25.9 &                25 & 23.0 & 18,19,20  \\
 80 &            A2537 & 347.106 &  --2.206 &  RELICS &  0.55 & 26.2 &                25 & 22.9 & 1  \\
 81 & MACSJ0454.1+0255 &  73.559 &   2.940 &  RELICS &  0.55 & 26.6 &                25 & 23.5 & 18,19,63  \\
 82 & MACSJ0232.2--4420 &  38.052 & --44.373 &  RELICS &  0.55 & 26.2 &                25 & 23.2 & 16,18,19  \\
 83 & MACSJ1131.8--1955 & 173.004 & --19.919 &  RELICS &  0.55 & 26.3 &                25 & 23.0 & 16,18,19,32  \\
 84 & MACSJ0308.9+2645 &  47.215 &  26.734 &  RELICS &  0.55 & 26.3 &                25 & 22.8 & 18,19  \\
 85 & MACSJ0257.1--2325 &  44.267 & --23.452 &  RELICS &  0.55 & 26.7 &                25 & 23.3 & 1  \\
 86 & MACSJ0600.1--2008 &  90.067 & --20.123 &  RELICS &  0.55 & 26.4 &                25 & 23.0 & 18,19,45  \\
 87 & MACSJ0358.8--2955 &  59.688 & --29.943 &  RELICS &  0.55 & 26.4 &                25 & 23.3 & 18,19  \\
 88 &         MACS0025 &   6.379 & --12.382 &  RELICS &  0.53 & 26.4 &                25 & 23.6 & 1  \\
 89 & MACSJ2211.7--0349 & 332.924 &  --3.858 &  RELICS &  0.53 & 26.2 &                25 & 23.1 & 16,18,19,45  \\
 90 & MACSJ0949.8+1708 & 147.445 &  17.122 &  RELICS &  0.53 & 26.0 &                25 & 23.0 & 18,19  \\
 91 & MACSJ0312.9+0822 &  48.220 &   8.352 &  RELICS &  0.53 & 26.0 &                25 & 23.4 & 16,18,19  \\
 92 &        0254--5857 &  43.518 & --58.963 &  RELICS &  0.53 & 26.5 &                25 & 23.2 & 16,18,19  \\
 93 & MACSJ0138.0--2155 &  24.498 & --21.951 & \nodata &  0.45 & 26.3 &                63 & 22.7 & 40,64  \\
 94 &        0151--5954 &  27.789 & --59.905 & \nodata &  0.39 & 26.6 &                64 & 22.8 & 16,30  \\
 95 &            34770 & 182.332 &  26.688 & \nodata &  0.34 & 26.4 &                65 & 23.3 & 65,66  \\
 96 &            34630 & 174.566 &  27.936 & \nodata &  0.34 & 26.5 &                65 & 22.6 & 66,67  \\
 97 & MACSJ2243.3--0935 & 340.831 &  --9.624 & \nodata &  0.34 & 26.3 &                65 & 22.9 & 45,67  \\
 98 & RXCJ2043.2--2144 & 310.826 & --21.712 & \nodata &  0.34 & 26.0 &                61 & 22.6 & 45  \\
 99 & MACSJ0150.3--1005 &  27.577 & --10.122 & \nodata &  0.33 & 26.3 &                66 & 22.9 & 16,40,68  \\
100 & MACSJ1717.1+2931 & 259.254 &  29.498 & \nodata &  0.33 & 26.3 &                67 & 22.4 & 40  \\
101 &            26029 & 215.233 &  39.935 & \nodata &  0.31 & 26.3 &                65 & 23.0 & 24,66,67  \\
\enddata
\tablecomments{This table is sorted by the decreasing order of total scientific integration time of archival WFC3-IR/F160W data.}
\vspace{-1.5mm}
\tablenotetext{a}{Names of \hst\ Treasury Program.}
\vspace{-1.5mm}
\tablenotetext{b}{Total scientific integration time. {For each HFF/BUFFALO cluster, the $t_\mathrm{obs}$ before slash stands for all observations (i.e., both original HFF and BUFFALO), and $t_\mathrm{obs}$ after slash stands for BUFFALO-only observations (used for source extraction, see Section~\ref{ss:02b_data}) that were split into four separate pointings.
}
}
\vspace{-1.5mm}
\tablenotetext{c}{$5\sigma$ {point-source} depth in the F160W band measured with $r=0\farcs4$ aperture. The depths of HFF clusters are calculated using BUFFALO-only data.}
\vspace{-1.5mm}
\tablenotetext{d}{\hst\ observation programs: (1) 12068 (Postman), (2) 13504 (Lotz), (3) 13790 (Rodney), (4) 14041 (Kelly), (5) 14199 (Kelly), (6) 14528 (Kelly), (7) 14872 (Kelly), (8) 15117 (Steinhardt), (9) 15308 (Gonzalez), (10) 11591 (Kneib), (11) 14038 (Lotz), (12) 14216 (Kirshner), (13) 13386 (Rodney), (14) 13495 (Lotz), (15) 12459 (Postman), (16) 13496 (Lotz), (17) 12103 (Postman), (18) 13498 (Lotz), (19) 12458 (Postman), (20) 14037 (Lotz), (21) 11099 (Bradac), (22) 13677 (Perlmutter), (23) 14327 (Perlmutter), (24) 15294 (Wilson), (25) 14096 (Coe), (26) 12790 (Postman), (27) 13177 (Bradac), (28) 12104 (Postman), (29) 15920 (Salmon), (30) 15883 (Schrabback), (31) 12102 (Postman), (32) 12461 (Riess), (33) 12457 (Postman), (34) 12065 (Postman), (35) 12360 (Perlmutter), (36) 12455 (Postman), (37) 12101 (Postman), (38) 15435 (Chisholm), (39) 15696 (Carton), (40) 12791 (Postman), (41) 12066 (Postman), (42) 12452 (Postman), (43) 12787 (Postman), (44) 12100 (Postman), (45) 12789 (Postman), (46) 12451 (Postman), (47) 12069 (Postman), (48) 12460 (Postman), (49) 12067 (Postman), (50) 12454 (Postman), (51) 11802 (Ford), (52) 12453 (Postman), (53) 12456 (Postman), (54) 12788 (Postman), (55) 11143 (Baker), (56) 13666 (Bradac), (57) 15831 (Bayliss), (58) 12197 (Richard), (59) 13767 (Trenti), (60) 14497 (Smit), (61) 14148 (Egami), (62) 12817 (Massey), (63) 14496 (Newman), (64) 14896 (Bayliss), (65) 13003 (Gladders), (66) 14205 (Newman), (67) 15670 (Boehringer).}
\vspace{-1.5mm}
\tablenotetext{e}{$5\sigma$ depth in the CH2 band measured with $r=1\farcs8$ aperture.}
\vspace{-1.5mm}
\tablenotetext{f}{\spitzer/IRAC observation programs: (1) 60034 (Egami), (2) 90009 (Bradac), (3) 90260 (Soifer), (4) 64 (Fazio), (5) 137 (Fazio), (6) 10171 (Soifer), (7) 83 (Rieke), (8) 90257 (Soifer), (9) 80168 (Bouwens), (10) 90258 (Soifer), (11) 40652 (Kocevski), (12) 90259 (Soifer), (13) 10170 (Soifer), (14) 3550 (Jones), (15) 40593 (Gonzalez), (16) 14253 (Stefanon), (17) 70053 (Brodwin), (18) 12005 (Bradac), (19) 12123 (Soifer), (20) 13165 (Bradac), (21) 20512 (Fadda), (22) 50393 (Kocevski), (23) 17 (Fazio), (24) 10043 (Sheth), (25) 14017 (Bradac), (26) 20740 (Holden), (27) 50726 (Holden), (28) 70063 (Holden), (29) 13210 (Bradac), (30) 80012 (Brodwin), (31) 60099 (Brodwin), (32) 14242 (Stroe), (33) 30659 (O'Dea), (34) 90233 (Lawrence), (35) 12030 (Strazzullo), (36) 70149 (Menanteau), (37) 545 (Egami), (38) 90213 (Bouwens), (39) 10140 (Coe), (40) 12095 (Egami), (41) 40204 (Kennicutt), (42) 11080 (Gonzalez), (43) 50610 (Yun), (44) 10015 (Johnston), (45) 90218 (Egami), (46) 10041 (Colbert), (47) 20439 (Egami), (48) 50096 (Martini), (49) 80162 (Lawrence), (50) 14304 (Stefanon), (51) 11121 (Finkelstein), (52) 40058 (Rines), (53) 80134 (Colbert), (54) 40311 (Lacy), (55) 20754 (Ellingson), (56) 60095 (Gal), (57) 60139 (Richards), (58) 10098 (Stern), (59) 14281 (Bradac), (60) 50138 (Rieke), (61) 3163 (Strauss), (62) 14130 (Bouwens), (63) 3720 (Huang), (64) 12127 (Newman), (65) 13006 (Trilling), (66) 70154 (Gladders), (67) 90232 (Rigby), (68) 12003 (Newman).}
\end{deluxetable*}

% \end{longrotatetable}
% \endlongtable

\startlongtable 

\begin{deluxetable*}{lrrrrrrrr} %[p!]
\tablecaption{Summary of measurements in the \hst/WFC3-IR F160W and \spitzer/IRAC CH1/CH2 bands \label{tab:02_phot}}
% \tablenum{01}
\tablewidth{0pt}
\tabletypesize{\scriptsize}
\tablehead{
\colhead{ID} & \multicolumn2c{Coordinates} & \colhead{Cluster Name} & \colhead{\hst/WFC3} & \multicolumn3c{\spitzer/IRAC} & \colhead{$\mu$\tablenotemark{$\star$}}  \\
\cline{2-3} \cline{6-8} %\vspace{-2mm}
\colhead{} & \colhead{RA} & \colhead{Dec}  & \colhead{} & \colhead{F160W\tablenotemark{$\dag$}} & \colhead{CH1} & \colhead{CH2} & \colhead{$R_\mathrm{e, circ}$\tablenotemark{$\ddag$}} & \colhead{} \\
\colhead{} & \colhead{(deg)} & \colhead{(deg)}  & \colhead{} & \colhead{(mag)} & \colhead{(mag)} & \colhead{(mag)} & \colhead{(\arcsec)} & \colhead{}
}
\startdata
ES-001$^{\vee}$   &   3.55824 & $-30.35491$ &             A2744 & $>$26.1 & 22.77$\pm$0.12 & 22.46$\pm$0.09 &    \nodata &     2.5 \\
ES-002$^{\vee}$   &   3.56268 & $-30.39096$ &             A2744 & $>$26.4 & 23.00$\pm$0.14 & 22.96$\pm$0.13 &    \nodata &     3.2 \\
ES-003$^{\vee}$   &   3.57555 & $-30.42436$ &             A2744 & $>$26.4 & 22.67$\pm$0.05 & 22.16$\pm$0.03 &    \nodata &     1.6 \\
ES-004$^{\vee}$   &   3.58133 & $-30.38023$ &             A2744 & 27.4$\pm$0.1 & 22.71$\pm$0.03 & 22.25$\pm$0.02 & 0.37$\pm$0.02 &     3.3 \\
ES-005$^{\wedge}$ &   3.62774 & $-30.39430$ &             A2744 & $>$26.2 & 23.78$\pm$0.12 & 23.09$\pm$0.06 &    \nodata &     1.4 \\
ES-006       &   8.03734 & $ 18.14496$ &  MACSJ0032.1+1808 & $>$26.6 & 22.70$\pm$0.03 & 22.32$\pm$0.03 & 0.43$\pm$0.17 &     2.9 \\
ES-007       &   8.05115 & $ 18.14700$ &  MACSJ0032.1+1808 & $>$26.5 & 23.81$\pm$0.68 & 22.95$\pm$0.27 & 0.79$\pm$0.08 &     3.2 \\
ES-008       &   8.05821 & $ 18.14266$ &  MACSJ0032.1+1808 & $>$26.6 & 22.92$\pm$0.26 & 22.41$\pm$0.15 & 0.54$\pm$0.07 &     4.2 \\
ES-009$^{\wedge}$ &  10.85795 & $-20.60537$ &             A2813 & $>$26.6 & 21.29$\pm$0.03 & 20.48$\pm$0.03 & 0.51$\pm$0.04 &     3.3 \\
ES-010       &  24.52332 & $-21.92225$ & MACSJ0138.0--2155 & $>$26.0 & 21.89$\pm$0.06 & 21.71$\pm$0.06 &    \nodata & \nodata \\
ES-011       &  28.17801 & $-13.95085$ &   CLJ0152.7--1357 & $>$27.3 & 23.56$\pm$0.04 & 23.41$\pm$0.04 &    \nodata &     2.4 \\
ES-012$^{\vee}$   &  39.96179 & $ -1.60007$ &             A370 & $>$26.3 & 23.25$\pm$0.05 & 22.82$\pm$0.04 &    \nodata &     2.3 \\
ES-013       &  44.28355 & $-23.45520$ & MACSJ0257.1--2325 & $>$26.7 & 23.55$\pm$0.07 & 23.28$\pm$0.09 & 0.22$\pm$0.08 &     1.0 \\
ES-014       &  44.30528 & $-23.42594$ & MACSJ0257.1--2325 & $>$26.6 & 25.08$\pm$0.75 & 23.64$\pm$0.10 &    \nodata &     1.2 \\
ES-015       &  59.69519 & $-29.92799$ & MACSJ0358.8--2955 & $>$26.3 & 23.29$\pm$0.05 & 23.08$\pm$0.05 & 0.70$\pm$0.16 &     2.3 \\
ES-016       &  64.04326 & $-24.11303$ &          MACS0416 & $>$26.5 & 23.49$\pm$0.06 & 22.97$\pm$0.04 & 0.55$\pm$0.12 &     1.5 \\
ES-017       &  64.04380 & $-24.08031$ &          MACS0416 & 27.2$\pm$0.2 & 23.25$\pm$0.33 & 22.71$\pm$0.22 &    \nodata &     2.7 \\
ES-018       &  64.39853 & $-11.91482$ & MACSJ0417.5--1154 & $>$26.7 & 23.05$\pm$0.11 & 22.36$\pm$0.04 &    \nodata &     1.9 \\
ES-019       &  64.40427 & $-11.90550$ & MACSJ0417.5--1154 & $>$26.7 & 23.30$\pm$0.06 & 22.55$\pm$0.04 &    \nodata &     2.7 \\
ES-020       &  67.38766 & $ -2.86797$ &          MACS0429 & $>$26.5 & 23.21$\pm$0.10 & 22.49$\pm$0.05 &    \nodata &     2.1 \\
ES-021       &  73.51120 & $  2.88566$ &  MACSJ0454.1+0255 & $>$26.6 & 23.19$\pm$0.16 & 22.35$\pm$0.05 &    \nodata & \nodata \\
ES-022$^{\vee}$   &  73.53382 & $-10.21773$ &             A521 & $>$27.0 & 24.98$\pm$0.28 & 23.76$\pm$0.14 &    \nodata & \nodata \\
ES-023       &  88.34406 & $-33.72470$ & MACSJ0553.4--3342 & $>$26.8 & 23.73$\pm$0.19 & 22.94$\pm$0.07 &    \nodata &     2.6 \\
ES-024$^{\vee}$   & 109.40894 & $ 37.77138$ &          MACS0717 & $>$26.3 & 22.45$\pm$0.04 & 21.96$\pm$0.03 &    \nodata &     2.2 \\
ES-025       & 110.85850 & $-73.44440$ & MACSJ0723.3--7327 & $>$27.0 & 22.81$\pm$0.07 & 22.35$\pm$0.04 &    \nodata &     2.2 \\
ES-026       & 116.21220 & $ 39.43703$ &          MACS0744 & $>$26.6 & 24.46$\pm$0.15 & 24.09$\pm$0.06 &    \nodata &     1.4 \\
ES-027       & 116.22664 & $ 39.44592$ &          MACS0744 & $>$26.9 & 22.10$\pm$0.11 & 21.57$\pm$0.07 & 0.29$\pm$0.15 &     2.8 \\
ES-028$^{\wedge}$ & 130.73140 & $ 36.36817$ &             A697 & $>$26.4 & 21.59$\pm$0.03 & 21.01$\pm$0.02 &    \nodata &     8.9 \\
ES-029$^{\wedge}$ & 130.76900 & $ 36.36602$ &             A697 & $>$26.0 & 22.78$\pm$0.05 & 22.21$\pm$0.03 &    \nodata &     2.6 \\
ES-030       & 137.82094 & $ 17.78358$ &  MACSJ0911.2+1746 & $>$26.0 & 22.77$\pm$0.02 & 22.49$\pm$0.02 & 0.29$\pm$0.09 &     1.2 \\
ES-031       & 145.22536 & $  7.72584$ &  MACSJ0940.9+0744 & $>$26.8 &        $>$23.3 & 21.75$\pm$0.08 & 0.38$\pm$0.23 & \nodata \\
ES-032       & 145.23096 & $  7.75463$ &  MACSJ0940.9+0744 & $>$26.8 & 22.99$\pm$0.26 & 22.26$\pm$0.13 &    \nodata & \nodata \\
ES-033       & 172.98836 & $-19.91272$ & MACSJ1131.8--1955 & $>$26.6 & 22.62$\pm$0.04 & 21.93$\pm$0.03 &    \nodata & \nodata \\
ES-034       & 172.98894 & $-19.91319$ & MACSJ1131.8--1955 & $>$26.6 & 22.62$\pm$0.03 & 21.97$\pm$0.03 &    \nodata & \nodata \\
ES-035       & 177.34485 & $ 22.40202$ &          MACS1149 & $>$26.1 & 22.49$\pm$0.04 & 22.11$\pm$0.03 &    \nodata &     1.3 \\
ES-036       & 177.36494 & $ 22.39943$ &          MACS1149 & $>$26.4 &        $>$22.5 & 23.22$\pm$0.62 &    \nodata &     1.3 \\
ES-037       & 177.37796 & $ 22.35166$ &          MACS1149 & $>$26.3 & 24.19$\pm$0.11 & 23.80$\pm$0.11 &    \nodata &     1.1 \\
ES-038$^{\vee}$   & 181.53986 & $ -8.82001$ &          MACS1206 & $>$26.3 & 23.39$\pm$0.11 & 23.08$\pm$0.06 & 0.51$\pm$0.14 &     1.5 \\
ES-039$^{\wedge}$ & 203.19160 & $ 50.55703$ &             A1758 & $>$26.6 & 22.51$\pm$0.14 & 21.93$\pm$0.07 &    \nodata &     1.8 \\
ES-040$^{\vee}$   & 203.20486 & $ 50.54511$ &             A1758 & $>$26.7 & 23.56$\pm$0.12 & 23.09$\pm$0.07 &    \nodata &     2.0 \\
ES-041       & 228.73633 & $-15.39844$ & MACSJ1514.9--1523 & $>$26.0 & 22.96$\pm$0.41 & 22.36$\pm$0.22 &    \nodata & \nodata \\
ES-042       & 228.75739 & $-15.38939$ & MACSJ1514.9--1523 & $>$26.5 & 22.71$\pm$0.13 & 22.49$\pm$0.09 &    \nodata & \nodata \\
ES-043       & 292.95038 & $-26.57872$ &          MACS1931 & $>$26.4 & 20.88$\pm$0.05 & 20.52$\pm$0.04 & 0.45$\pm$0.06 &     6.8 \\
ES-044       & 310.81604 & $-21.74442$ & RXCJ2043.2--2144 & $>$26.2 & 23.44$\pm$0.36 & 22.22$\pm$0.14 &    \nodata & \nodata \\
ES-045$^{\vee}$   & 322.33627 & $ -7.69648$ &          MACS2129 & $>$26.4 & 24.24$\pm$0.25 & 23.64$\pm$0.14 &    \nodata &     2.4 \\
ES-046$^{\vee}$   & 322.36267 & $ -7.71412$ &          MACS2129 & $>$26.3 & 24.51$\pm$0.23 & 23.87$\pm$0.11 &    \nodata &     1.4 \\
ES-047$^{\vee}$   & 325.08832 & $-23.65191$ &            MS2137 & $>$26.3 & 22.62$\pm$0.06 & 22.05$\pm$0.04 &    \nodata &     1.4 \\
ES-048       & 340.83810 & $ -9.58147$ & MACSJ2243.3--0935 & $>$26.1 & 23.20$\pm$0.15 & 22.81$\pm$0.14 &    \nodata & \nodata \\
ES-049$^{\vee}$   & 342.15656 & $-44.56572$ &            AS1063 & $>$26.1 & 22.34$\pm$0.03 & 22.08$\pm$0.03 & 0.38$\pm$0.13 &     1.3 \\
ES-050       & 342.16800 & $-44.48981$ &            AS1063 & $>$25.7 & 21.47$\pm$0.03 & 21.16$\pm$0.02 &    \nodata &     1.1 \\
ES-051$^{\vee}$   & 342.22214 & $-44.57252$ &            AS1063 & $>$26.0 & 23.67$\pm$0.07 & 23.27$\pm$0.06 &    \nodata &     1.1 \\
ES-052$^{\vee}$   & 347.09330 & $ -2.21299$ &             A2537 & $>$26.2 & 22.68$\pm$0.06 & 22.20$\pm$0.03 &    \nodata &     2.3 \\
ES-053$^{\vee}$   & 347.11108 & $ -2.17515$ &             A2537 & $>$26.0 & 23.26$\pm$0.07 & 22.60$\pm$0.05 &    \nodata &     1.9 \\
\enddata
\tablecomments{ES-006/07/08, ES-018/19, ES-028/29 and ES-045/46 are multiply imaged systems.
{ES-036 has a large uncertainty of CH2 magnitude ($23.22\pm0.62$) because of a steep background brightness slope (and therefore large uncertainty) contributed by neighboring bright sources. The magnitude modeled with GALFIT is $23.45\pm0.16$.}}
\vspace{-1.5mm}
\tablenotetext{\dag}{$5\sigma$ limit of F160W magnitude {assuming point-source profile. If the source is extended, the actual magnitude limit could be higher by $\sim$0.5\,mag (assuming $R_\mathrm{e,circ} = 0\farcs25$, $b/a=0.4$) to 1.2\,mag (assuming $R_\mathrm{e,circ} = 0\farcs45$, $b/a=0.2$; see Section 3.1).}}
\vspace{-1.5mm}
\tablenotetext{\ddag}{Circularized effective radius measured in the IRAC CH2 band with \textsc{galfit} (Section~\ref{ss:03b_galfit}).}
\vspace{-1.5mm}
\tablenotetext{\star}{Lensing magnification factor (Section~\ref{ss:04b_lensing}).} 
\vspace{-1.5mm}
\tablenotetext{\vee}{Undetected ({$<4\sigma$}) in the MIPS 24\,\micron\ band. The typical $4\sigma$ upper limit is 61\,\si{\micro Jy}.}
\vspace{-1.5mm}
\tablenotetext{\wedge}{Detected ({$>4\sigma$}) in the MIPS 24\,\micron\ band. 
The measured MIPS 24\,\micron\ flux densities are 46$\pm$8\,\si{\micro Jy} (ES-005), 279$\pm$29\,\si{\micro Jy} (ES-009) 263$\pm$11\,\si{\micro Jy} (ES-028),  134$\pm$15\,\si{\micro Jy} (ES-029) and 67$\pm$11\,\si{\micro Jy} (ES-039).
}
\end{deluxetable*}

\begin{deluxetable*}{@{\extracolsep{10pt}}rrrrrr} %[p!]
\tablecaption{Properties of $H$-faint galaxies \label{tab:03_stack}}
% \tablenum{01}
\tablewidth{0pt}
\tabletypesize{\footnotesize}
\tablehead{
\colhead{} & \colhead{Individual} & \multicolumn{3}{c}{Stacked\tablenotemark{a} } \\ 
\cline{2-2}\cline{3-5}
\colhead{Parameters} & \colhead{ES-009} &\colhead{All (52)}  & \colhead{MIPS-faint (16)} & \colhead{MIPS-bright (4)}  
}
% \decimalcolnumbers
\startdata 
& \multicolumn{4}{c}{Photometric Properties\tablenotemark{b}} \\ 
  \cline{2-5}
      WFC3-IR/F105W (mag) &        $>$26.3 &        $>$28.7 &        $>$28.1 &        $>$27.5 \\
      WFC3-IR/F110W (mag) &        \nodata &        $>$28.2 &        $>$27.0 &        $>$26.1 \\
      WFC3-IR/F125W (mag) &        $>$25.6 &        $>$28.5 &        $>$27.9 &        $>$26.1 \\
      WFC3-IR/F140W (mag) &        $>$25.7 &        $>$27.0 &        $>$25.8 &        $>$26.7 \\
      WFC3-IR/F160W (mag) &        $>$26.2 & 26.79$\pm$0.21 & 26.38$\pm$0.19 & 26.52$\pm$0.23 \\
           IRAC/CH1 (mag) & 21.29$\pm$0.03 & 22.95$\pm$0.11 & 23.25$\pm$0.06 & 22.79$\pm$0.07 \\
           IRAC/CH2 (mag) & 20.48$\pm$0.03 & 22.61$\pm$0.03 & 22.77$\pm$0.02 & 22.14$\pm$0.04 \\
           IRAC/CH3 (mag) &        \nodata & 22.31$\pm$0.16 & 22.92$\pm$0.18 & 21.78$\pm$0.14 \\
           IRAC/CH4 (mag) &        \nodata & 22.07$\pm$0.18 & 22.34$\pm$0.32 & 21.86$\pm$0.16 \\
 MIPS  24\,\micron\ (mJy) &  0.28$\pm$0.03 &  0.03$\pm$0.02 &        $<$0.02 &  0.13$\pm$0.01 \\
 PACS 100\,\micron\ (mJy) &    4.0$\pm$0.5 &    0.4$\pm$0.2 &         $<$0.7 &    1.0$\pm$0.2 \\
 PACS 160\,\micron\ (mJy) &   12.6$\pm$1.0 &    0.8$\pm$0.4 &         $<$1.4 &    2.3$\pm$0.8 \\
SPIRE 250\,\micron\ (mJy) &   37.6$\pm$5.4 &    4.0$\pm$1.0 &         $<$4.8 &    7.3$\pm$3.6 \\
SPIRE 350\,\micron\ (mJy) &   51.5$\pm$4.2 &    4.7$\pm$1.1 &         $<$4.7 &    7.1$\pm$3.3 \\
SPIRE 500\,\micron\ (mJy) &   46.6$\pm$4.4 &    5.9$\pm$2.0 &    7.6$\pm$3.0 &        $<$12.3 \\
\cline{2-5}
& \multicolumn{4}{c}{Physical Properties\tablenotemark{c}} \\ 
  \cline{2-5}
$z_\mathrm{EAZY}$ & \nodata & $3.9_{-0.4}^{+0.4}$ & $3.6_{-0.3}^{+0.4}$ & $2.6_{-0.4}^{+0.4}$ \\
$z_\mathrm{MAGPHYS}$ & $3.2_{-0.1}^{+0.1}$ & $2.7_{-0.4}^{+1.0}$ & $3.2_{-0.6}^{+0.6}$ & $1.8_{-0.2}^{+0.3}$ \\
$\mu$ & 3.3  & 2.2  & 2.0  & 2.2  \\
$A_V$ & $4.5\pm0.3$ & $2.6\pm0.3$ & $2.3\pm0.3$ & $4.0\pm0.6$ \\
$\log[M_\mathrm{star}/(\mathrm{M}_\odot)]$ & $11.2\pm0.2$ & $10.3\pm0.2$ & $10.1\pm0.2$ & $10.2\pm0.2$ \\
$\log[L_\mathrm{IR}/(\mathrm{L}_\odot)]$ & $12.8\pm0.2$ & $12.0\pm0.2$ & $11.8\pm0.2$ & $11.9\pm0.3$ \\
$\log[\mathrm{SFR}/(\mathrm{M}_\odot\,\mathrm{yr}^{-1})]$ & $2.7\pm0.2$ & $2.0\pm0.2$ & $1.7\pm0.2$ & $1.9\pm0.3$ \\
\enddata
\tablenotetext{a}{From left to right: stack of all the 52 sources except for ES-009; 16 sources which are undetected at 24\,\micron; four sources (ES-005, ES-028, ES-029 and ES-039) which are detected at 24\,\micron. }
\vspace{-1.5mm}
\tablenotetext{b}{Normalized by the median IRAC/CH2 flux densities (Section~\ref{ss:03d_stack}). Upper limits are at $3\sigma$.}
\vspace{-1.5mm}
\tablenotetext{c}{$A_V$, $M_\mathrm{star}$, $L_\mathrm{IR}$ and SFR are derived with \textsc{cigale} (Section~\ref{ss:04c_phys}). 
Lensing magnification is corrected for $M_\mathrm{star}$, $L_\mathrm{IR}$ and SFR.}
\end{deluxetable*}

% \input{table/tb04_irphot}

%% For this sample we use BibTeX plus aasjournals.bst to generate the
%% the bibliography. The sample63.bib file was populated from ADS. To
%% get the citations to show in the compiled file do the following:
%%
%% pdflatex sample63.tex
%% bibtext sample63
%% pdflatex sample63.tex
%% pdflatex sample63.tex

\bibliography{00_main}{}
\bibliographystyle{aasjournal}

%% This command is needed to show the entire author+affiliation list when
%% the collaboration and author truncation commands are used.  It has to
%% go at the end of the manuscript.
%\allauthors

%% Include this line if you are using the \added, \replaced, \deleted
%% commands to see a summary list of all changes at the end of the article.
%\listofchanges

\end{document}